\documentclass[a4paper,fleqn]{cas-sc}

\usepackage{float}
\usepackage{subcaption}
\usepackage{placeins}
\usepackage{pifont}
\usepackage{longtable}
\usepackage{enumitem}
\usepackage{siunitx}
\usepackage[authoryear]{natbib}

\def\tsc#1{\csdef{#1}{\textsc{\lowercase{#1}}\xspace}}
\tsc{WGM}
\tsc{QE}

\begin{document}
\let\WriteBookmarks\relax
\def\floatpagepagefraction{1}
\def\textpagefraction{.001}
\let\printorcid\relax 


\shortauthors{Ding et al.}

\title[mode = title]{URBAN-SPIN: A street-level bikeability index to inform design implementations in historical city centres}

\author[1,2]{Haining Ding}[style=chinese]

\cormark[1] 
\credit{}

\author[1,2]{Chenxi Wang}

\credit{}

\author[1,2]{Simon Ladouce}

\credit{}
\author[1,2]{Michal Gath-Morad}
\cormark[1] 

\credit{}

\address[1]{Cambridge Cognitive Architecture, Department of Architecture, University of Cambridge, UK}
\address[2]{NeuroCivitas Lab for NeuroArchitecture, Centre for Research in the Arts, Social Sciences and Humanities (CRASSH), University of Cambridge, UK}

\cortext[1]{Corresponding author. Cambridge Cognitive Architecture, Department of Architecture, University of Cambridge, UK. \\
\hspace*{2em} \textit{Email Address}: hd518@cam.ac.uk (H. Ding), mg2068@cam.ac.uk (M. Gath-Morad).}

\begin{abstract}
Cycling is reported by an average of 35\% of adults at least once per week across 28 countries and is increasingly positioned as an everyday mode of urban mobility relevant to decarbonisation and public health. Yet as vulnerable road users exposed to surroundings, cyclists experience the street intensely, making exposure to risk and discomfort immediate concerns for cycling uptake. While bikeability research has expanded, street-level features shaping perceived cycling experience remain under-analysed, particularly in how their effects shift or cumulate across street typologies. To address this gap, this study develops URBAN-SPIN, a data-driven framework for evaluating how visual and spatial configurations shape perceived bikeability across street typologies. Using first-person cycling videos from 116 street segments in Cambridge's historical city centre, where spatial constraints limit large-scale infrastructural change, we extract fine-grained streetscape indicators through computer vision and pair them with built-environment variables and survey ratings from 75 participants to construct a segment-level Bikeability Index. Statistical analysis shows that perceived bikeability arises from cumulative, context-specific interactions among environmental features. Greenness showed the strongest and most stable associations with cycling experience, while openness was positively associated with safety and comfort. Enclosure showed non-linear associations, and imageability and building continuity varied by typology. Model-informed visual scenarios illustrate how typology-specific street-level adjustments could shift predicted perceptual outcomes without large-scale structural interventions. The framework offers a transferable model for evaluating cycling conditions across urban street typologies, supporting perceptually attuned design strategies in historical contexts where interventions are constrained.\end{abstract}




\maketitle


\section{Introduction}

What makes a place feel bikeable? As cycling becomes increasingly embedded in sustainable mobility agendas, this question has shifted from normative advocacy to a demand for precise empirical articulation \citep{ahmed_assessing_2024, cunha_assessing_2023, rerat_build_2024, liu_urban_2022}. Over the past two decades, bikeability has therefore evolved into a critical research domain, bridging urban design, transportation planning, and environmental psychology \citep{nazemi_studying_2021, zuo_bikeway_2019, gotardi_affordance-based_2024} to understand how the built environment influences the likelihood and quality of cycling \citep{gao_accessibility_2024, kim_assessing_2025, dong_assessing_2023}.

Previous research affirms the broad benefits of urban cycling, highlighting its role in decarbonisation, public health, and spatial equity \citep{basil_exploring_2023, bogacz_modelling_2021, aldred_diversifying_2017}. Much of this work has examined bikeability through macro-scale measures such as network connectivity, land-use diversity, and proximity to destinations \citep{beura_development_2022, zhang_encouraging_2024, weikl_data-driven_2023, ciris_investigating_2024, rerat_build_2024}. These approaches are effective for city-level comparison and policy benchmarking, but often overlook the fine-grained conditions that shape how cycling is experienced on the street. Experiential and affective qualities, such as how safe or comfortable a street feels, remain difficult to capture when bikeability is represented mainly through infrastructure and accessibility metrics \citep{arellana_developing_2020, ito_assessing_2021, dai_assessing_2023}.

In contrast, a growing body of micro-scale research examines street conditions from the cyclist's perspective, showing that perceived bikeability can be shaped by cycling facilities and visible street features, such as vegetation or continuous building frontages \citep{ahmed_bicycle_2024, biassoni_choosing_2023, bialkova_how_2022, nielsen_bikeability_2018}. By emphasising the street as the immediate setting of cycling, these works link visible conditions to movement, affect, and willingness to cycle \citep{hankey_predicting_2021, gavriilidou_empirical_2021, guo_moderation_2024}. This link can be understood through affordance theory, which treats environmental features as action-relevant information that cyclists use to judge where movement feels possible, safe, or constrained \citep{gibson_ecological_1979, gotardi_affordance-based_2024}. This focus is especially relevant in historical urban contexts such as Cambridge, where centuries-old layouts constrain large-scale renovation. In such settings, incremental street-level adjustments become a critical scale for design intervention, while their perceptual effects require empirical assessment \citep{carse_factors_2013, aldred_outside_2010}. However, existing micro-scale studies often remain divided between objective measurements, such as correlations between street features and cycling volumes, and subjective accounts, such as self-reported safety or comfort \citep{yin_commute_2024, ni_evaluation_2024, hardinghaus_more_2021}. Rarely are these approaches integrated into a framework that accounts for both the physical characteristics of space and the perceptual responses they elicit. This separation limits the capacity of current research to model bikeability as an experience shaped by both physical street conditions and subjective perception.

Recent studies move beyond facility counts to explain how cycling is enabled and experienced through network structure, temporal dynamics, and visual context. \citet{smith_when_2025} show that continuous low-stress corridors shape commuter uptake, while citywide analyses use dockless-bike traces and street-view imagery to explain spatial and temporal variation in cycling environments \citep{zhou_exploring_2024, gao_pedaling_2025}. At the street level, protected-facility studies and semantic-segmentation workflows increasingly link visual elements to bikeability, perception proxies, and nonlinear micro-scale effects \citep{xu_protected_2025, xu_exploring_2026, lee_nonlinear_2025}. Perception modelling has also expanded through transfer learning, multimodal street-view analysis, and large-sample experiments on affective responses to urban morphology \citep{le_scalable_2025, zhang_urban_2025, qiu_subjective_2023, qi_how_2026, froehlich_streetreaderai_2025}. However, most workflows still infer perception from vehicle-mounted panoramas instead of cyclists' first-person views and often generalise features across settings, which limits segment-level and context-aware insight in historically constrained streets.

In addition to this methodological fragmentation, a further limitation lies in the insufficient attention to typological context. Current studies tend to treat street-level features as discrete elements abstracted from their spatial settings \citep{danish_citizen_2025, von_stulpnagel_matter_2024, sanchez-vaquerizo_virtual_2024, nikiforiadis_can_2019}. However, in practice, the perceptual and functional effects of these features are embedded within the street's morphological configuration. For instance, added greenery may enhance comfort on a leisurely greenway but hinder visibility on a narrow and high-traffic corridor. This typological contingency helps explain why the same feature may support ease and legibility in one setting while increasing ambiguity in another. Without considering such typology-sensitive and context-dependent interactions, design recommendations risk oversimplification and reduced effectiveness across diverse street types \citep{rui_beyond_2024, ma_measuring_2021}.

Despite growing recognition that street-level features and subjective impressions can enrich cycling studies, their translation into street-level design scenarios remains limited. While several recent studies have introduced data-driven frameworks to inform city-scale cycling policy and interventions \citep{pais_multicriteria_2022, cai_sidewalk-based_2024, mahfouz_road_2023}, few incorporate subjective perception or address street-level potentials. To the best of our knowledge, no existing study has systematically integrated perceptual data into typology-sensitive design strategies at the street level. As a result, bikeability research often remains abstracted from street-level implementation, leaving a gap between empirical analysis and design-oriented interpretation.

To address these gaps, this research investigates how micro-scale street features shape subjective cycling experience, and how such insights can inform street-level and typology-sensitive design scenarios. The study focuses on historical urban contexts, specifically the city centre of Cambridge, where large-scale infrastructural transformation is often infeasible, and even subtle street-level interventions may meaningfully enhance perceived bikeability. The research question is: How and to what extent do specific street-level features, within different street typologies, shape subjective cycling experiences in the historical city centre of Cambridge?

This study advances a replicable, perception-led framework to quantify how spatial and visual features jointly shape subjective cycling experience. Conceptually, the framework is informed by affordance theory, positioning street-level bikeability as an interaction between visible environmental conditions and cyclists' perceptual evaluation during movement \citep{gibson_ecological_1979, gotardi_affordance-based_2024}. Drawing on an emerging body of literature on bikeability \citep{ros-mcdonnell_development_2020, karolemeas_measure_2022, schmid-querg_munich_2021, hardinghaus_more_2021, krenn_development_2015, arellana_developing_2020, tran_cyclists_2020, codina_built_2022, porter_bikeability_2020, winters_bike_2016, lowry_assessment_2012, dai_assessing_2023, ito_assessing_2021, lin_assessing_2018}, it constructs a composite index grounded in subjective evaluations and develops illustrative, AI-assisted design scenarios derived from model outputs. Specifically, the paper contributes: (1) a typology-sensitive Bikeability Index at the street-segment scale; (2) an integrated pipeline that links machine vision, subjective experience, and built-environment indicators in a historical-city context; and (3) typology-tailored design scenarios that show how small, context-aware modifications could shift predicted comfort, safety, and legibility across distinct street types.

\section{Related works}

\subsection{Bikeability index}

The term \textit{bikeability} emerged as a parallel to \textit{walkability} \citep{kellstedt_scoping_2021, damiani_high-resolution_2021}, reflecting growing interest in how urban form facilitates active mobility beyond motorised transport. Initially framed within public health and planning discourse, it has gradually taken on a more technical meaning, encompassing spatial conditions and cyclists' subjective evaluations of those conditions \citep{castanon_bikeability_2021, hsu_hybrid_2023}. As the concept has become increasingly spatialised and multidimensional, composite indices have gained traction as a means to capture its complexity \citep{cao_multi-index_2024, nordengen_national_2021}. By translating environmental or experiential variables into evaluative scores, these indices offer structured frameworks for benchmarking and policy alignment across urban contexts. Yet each index reflects distinct priorities and inherits constraints from its chosen methods \citep{kellstedt_scoping_2021, ahmed_bicycle_2024, castanon_bikeability_2021}. To the best of our knowledge, existing indices rarely integrate subjective and objective measures at the street-segment scale in a way that supports typology-sensitive design interpretation (see Table \ref{tab2}).

Most indices follow a shared construction logic: selecting indicators, sourcing data, assigning weights, and aggregating outputs \citep{schmid-querg_munich_2021, winters_bike_2016}. Yet methodological choices reflect divergent priorities. GIS-based models prioritise accessibility, connectivity, or land-use mix through geometric overlays \citep{lin_assessing_2018, ros-mcdonnell_development_2020, karolemeas_measure_2022}, while perception-led approaches rely on surveys or on-site audits to capture user-reported comfort, safety, or visual appeal \citep{codina_built_2022, porter_bikeability_2020, arellana_developing_2020}. Recent advances in computer vision allow large-scale extraction of environmental features from street-level imagery \citep{ito_assessing_2021, hou_global_2024}, though these features remain largely detached from first-person cycling experience. Even when incorporating visible elements such as greenery or sky \citep{dai_assessing_2023}, few studies anchor them to first-person perceptual or behavioural contexts.

Weighting and aggregation methods further encode assumptions about how cycling conditions should be evaluated. Equal weighting offers statistical neutrality but may obscure contextual variation. By contrast, expert-assigned and preference-based weights allow differentiation by user type, trip purpose, or setting \citep{arellana_developing_2020, lin_assessing_2018, castanon_bikeability_2021}. Aggregation strategies range from additive models to machine-learning composites, each introducing distinct assumptions about indicator interaction \citep{schmid-querg_munich_2021, dai_assessing_2023}.

While \citet{ito_assessing_2021} and \citet{codina_built_2022} draw from all three analytical strands outlined in Table~\ref{tab2}, their citywide and neighbourhood-level scopes leave less room to identify how specific street-segment conditions shape cyclists' subjective experience. This limits their use for translating perceptual evidence into typology-sensitive design interpretation, particularly where historical morphology constrains conventional cycling infrastructure.

\begin{table}
\centering
\setlength{\tabcolsep}{10pt}
\renewcommand{\arraystretch}{2.5}
\footnotesize

\begin{tabular}{@{}p{2cm}p{1.5cm}p{1.5cm}p{2.8cm}p{2.8cm}p{2.8cm}@{}}
\toprule
\textbf{Publication} & \textbf{Urban Scale} & \textbf{Data Source} & \textbf{Machine-Vision Streetscape} & \textbf{Cycling Experience} & \textbf{Spatial/Built Environment} \\
\midrule
\cite{lowry_assessment_2012} & City scale & GIS layers, census data, parcel data & N/A & Level of Traffic Stress &  Network connectivity, facility access, land use mix, slope, topography, destination proximity \\
\cite{arellana_developing_2020} & City scale & Survey, GSV, GIS & N/A &Safety, comfort, preferences by user type & Cycle facility type, width, surface, traffic speed, volume, slope, intersection control, lighting, wayfinding, land use mix, density, connectivity \\
\cite{karolemeas_measure_2022} & City scale & GIS, transport plans & N/A & N/A & Bike lane continuity, intersection density, land use mix, road width, traffic speed, slope \\
\cite{dai_assessing_2023} & City scale & OSM Data, POIs data & N/A & N/A & Raster-based slope, elevation, land use, population density, road density, speed limits, facility access \\
\cite{ito_assessing_2021} & City scale & SVI, survey, OSM, DEM, AQI & greenery, buildings, water, presence of street elements &
attractiveness, spaciousness, cleanliness, building design attractiveness, safety as a cyclist, beauty, attractiveness for living & Road type and width, land use mix, slope, pavement type, transit facility density, vehicle presence, etc. \\
\cite{codina_built_2022} & Neighbourhood scale & Survey, on-site audit, GIS & Connectivity, greenness, imageability & Comfort, directness perception & Slope, land use mix, infrastructure proximity, crash frequency \\
\cite{porter_bikeability_2020} & Neighbourhood scale & GIS, GPS, survey & N/A & N/A & Land use mix, connectivity, facility access, road type, traffic volume \\
\cite{ros-mcdonnell_development_2020} & Regional scale & GIS, environmental datasets & N/A & N/A & Climate, slope, transport network density \\
\cite{lin_assessing_2018} & Regional scale & Road network, OSM, POIs & N/A & N/A & GANP-weighted road importance, land use, slope, intersections \\
\bottomrule
\end{tabular}
\vspace{5mm}
\caption{Review of Existing Bikeability Indices.}
\label{tab2}
\end{table}

\subsection{Bikeability index indicators}

In this research, the bikeability index integrates spatial and perceptual indicators that reflect the experiential and environmental dimensions of cycling. First-person video and computer vision are used because they make visible street conditions measurable from the cyclist's viewpoint and allow these conditions to be linked with subjective experience ratings. Indicator selection is informed by a comprehensive literature review of existing bikeability indices (see Table~\ref{tab2}) and guided by three questions: Which street-level factors are associated with cycling experience? Which are relevant to historically constrained street settings? Which indicators can be detected and measured through first-person video footage and computer vision techniques? Based on this review, the indicators are grouped into three functional categories: Machine-Vision Streetscape, Cycling Experience, and Spatial / Built-Environment Indicators, each outlined below. Full indicator definitions, formulas, and explanations are provided in Supplementary Material Section A, Table S1.

\subsubsection{Machine-vision streetscape}

Recent research shows that machine-vision streetscape can describe how cyclists encounter visible street form from a first-person viewpoint. First developed extensively in \textit{walkability} studies and now applied to cycling, this approach captures fine-grained variation in street-level form and experience. In this study, these indicators are used to measure visible environmental conditions through which surrounding street form may support or constrain cycling movement. This study uses six machine-vision indicators: openness, greenness, enclosure, non-motorised lane proportion, imageability, and building continuity. They were selected for empirical relevance to cycling, suitability to constrained historical settings, and detectability through semantic segmentation of first-person videos.

Openness broadly refers to the degree of visual and physical exposure within a space; in this study, it is operationalised through sky visibility, which approximates vertical clearance and perceived spaciousness in the cyclist's field of view and is positively associated with comfort and reduced travel stress \citep{tang_measuring_2019, li_using_2018}. Greenness denotes visible vegetation such as trees, grass, or planting strips along the street edge and has been linked to aesthetic quality and increased cycling likelihood \citep{gu_deep_2019, lu_associations_2019}. Enclosure reflects the extent to which adjacent vertical elements bound the cyclist's visual field; moderate enclosure enhances spatial definition and coherence, whereas excessive enclosure may trigger discomfort or perceived risk \citep{park_street_2019, ewing_streetscape_2016}.

Non-motorised lane proportion quantifies the share of visible road space dedicated to cyclists and pedestrians and serves as a proxy for perceived accessibility and protection from vehicular traffic \citep{rui_decoding_2024}. Imageability, following \citet{lynch_image_2008}, captures recognisable visual elements that support spatial orientation. Operationally, this study represents imageability through the normalised presence of person, bicycle, pole, traffic light, traffic sign, car, truck, bus, and motorcycle pixels in the field of view, treating these elements as proxies for recognisable street information that may support traffic awareness and spatial orientation. Building continuity measures facade coherence and uninterrupted street walls, contributing to visual rhythm and environmental legibility \citep{wu_using_2023, rui_beyond_2024}.

\subsubsection{Cycling experience}

Cycling experience is the subjective evaluation of a journey, encompassing affective responses and usability judgements elicited by the spatial environment. In this study, the focus is on momentary perception during cycling, where riders evaluate visible street conditions while moving through the environment. This perspective complements the machine-vision streetscape indicators by describing how cyclists respond to perceptual information during movement.

This study assesses cycling experience with five indicators: perceived quality, safety, comfort, pleasure, and arousal. The first three draw on established frameworks such as Perceived Level of Service (PLOS) and Quality of Service (QoS), commonly used in transport research to evaluate infrastructure satisfaction, perceived exposure to harm, and physical ease of use \citep{florian_decision_2024, magnana_implicit_2022, nikiforiadis_investigating_2023, liu_evaluating_2020}. These indicators provide structured measures of usability across varied street conditions.

To capture emotional states, the study uses the Affect Grid, a compact and validated tool that maps immediate responses along two axes: pleasure (valence) and arousal (activation) \citep{russell_affect_1989, zhou_can_2024}. It enables standardised quantification of affective states evoked by physical surroundings, from stress and fatigue to excitement and relaxation.

\subsubsection{Built environment}

Built environment indicators are externally measurable properties of urban form and function that shape cycling behaviour through spatial configuration, land-use intensity, and regulation. To stabilise statistical models and ensure comprehensive evaluation, we included seven features from prior studies as control variables. These indicators capture objective, spatially standardised aspects of urban form, including building density, intersection density, land-use mix (Shannon index), point-of-interest diversity (Simpson index), floor area ratio (FAR), slope, and traffic speed limits \citep{ito_assessing_2021, ito_examining_2024, rui_beyond_2024, peterson_mapping_2017}. All variables came from publicly available datasets such as OpenStreetMap (OSM), official land-use records, and digital elevation models.

These features account for macro-scale environmental variation across street segments and help isolate the influence of perceptual and experiential indicators. While not the focus of interpretation in this study, their inclusion supports analytical rigour and enables controlled comparison between spatial typologies.

\section{Data and methods}

\begin{figure}
    \centering
    \includegraphics[width= 1 \linewidth]{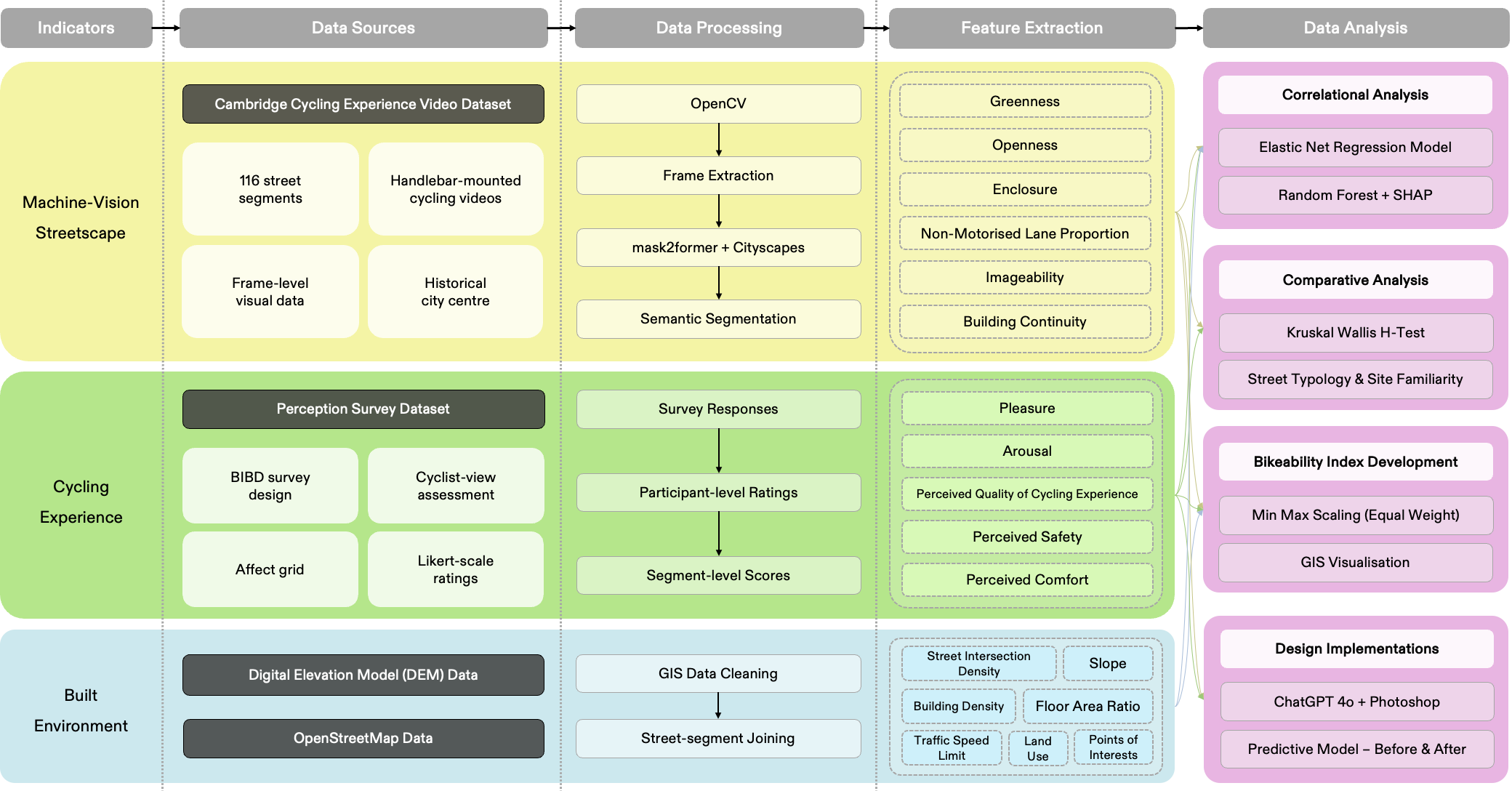} 
    \caption{Research methodological framework.
    \newline The methodological workflow links three data sources across 116 street segments. The Cambridge Cycling Experience Video Dataset (CCEVD) provides first-person videos recorded with a handlebar-mounted camera in real-world cycling conditions in Cambridge. The Balanced Incomplete Block Design (BIBD) perception survey captures five dimensions of subjective cycling experience: perceived quality, safety, comfort, pleasure and arousal. Built-environment indicators are derived from OpenStreetMap (OSM) and a digital elevation model (DEM).
}
    \label{3.0} 
\end{figure}

This study uses four data sources: the Cambridge Cycling Experience Video Dataset (CCEVD), survey responses, OpenStreetMap (OSM), and a Digital Elevation Model (DEM), to evaluate 18 indicators across three categories: machine-vision streetscape, cycling experience, and built environment. Using Cambridge as a case study, the research focuses on a high-density historical urban core to examine how micro-scale spatial features relate to perceptual responses under constrained morphological conditions. An expert interview was also conducted to assess the practical relevance of findings within planning and policy contexts. The composite Bikeability Index was developed through equal weighting across the three indicator domains and normalised for cross-segment comparison. Figure~\ref{3.0} presents the methodological framework.

\FloatBarrier

\subsection{Study design and data collection}

\subsubsection{Study area and street typologies}

\begin{figure} 
    \centering
    \includegraphics[width= 1 \linewidth]{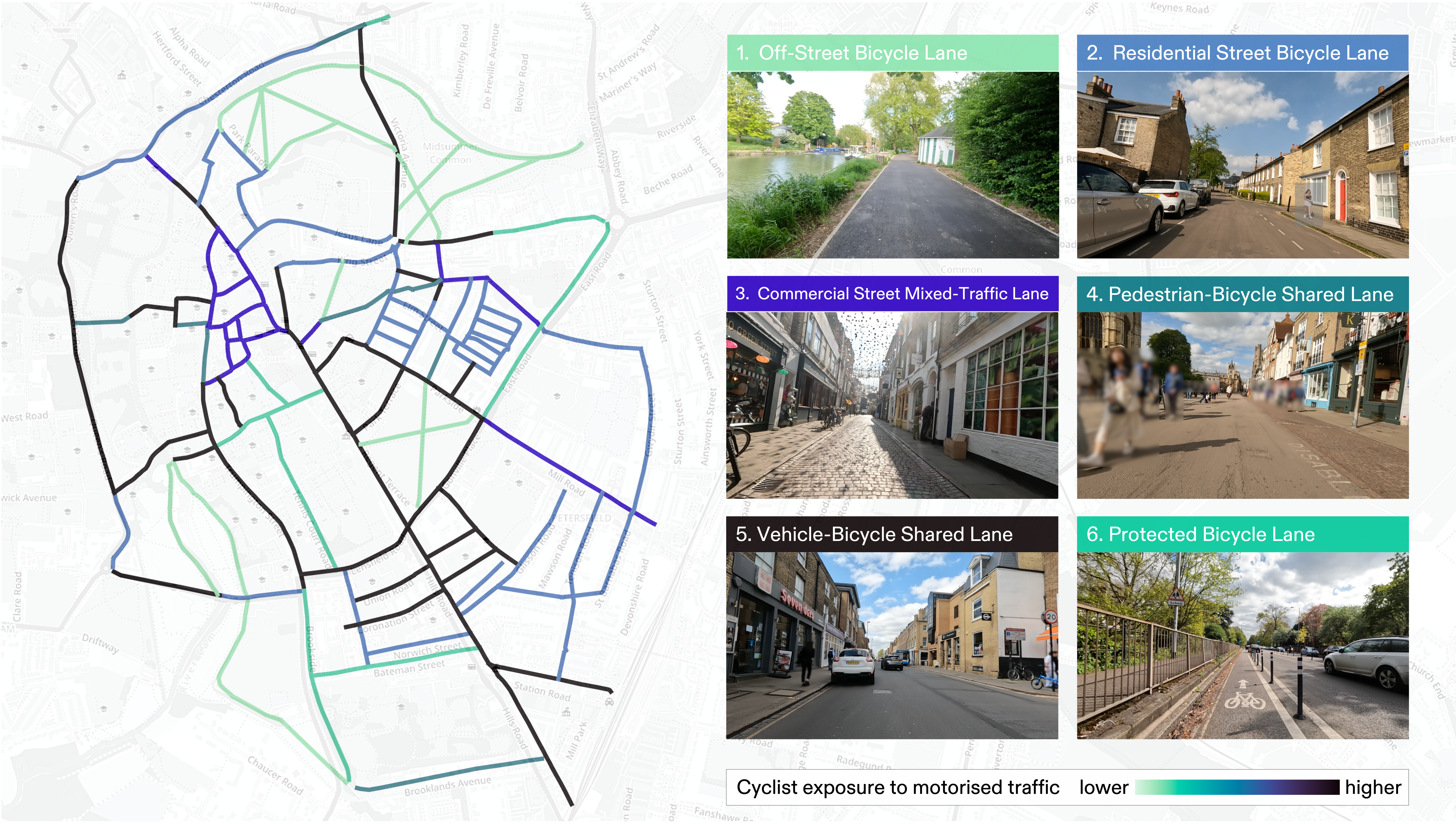} 
    \caption{Study area and street typologies.
    \newline To account for spatial heterogeneity, all segments are classified into six typologies adapted from established classifications \citep{houde_ride_2018, nazemi_studying_2021, noland_understanding_2023, teschke_proximity_2017, zhang_measuring_2022}: (1) off-street bicycle lanes, (2) residential street bicycle lanes, (3) commercial street mixed-traffic lanes, (4) pedestrian-bicycle shared lanes, (5) vehicle-bicycle shared lanes, and (6) protected bicycle lanes. Colours provide an indicative ordering of cyclist exposure to motorised traffic across the six street typologies, based on their modal configuration and observed street conditions, with darker tones representing greater exposure. Each typology represents a distinct configuration of spatial form, supporting comparative analysis of how identical features, such as greenness or enclosure, may operate differently across street types. This typological structure informs all subsequent modelling and design implementation stages.}
    \label{fig 2.1-2} 
\end{figure}

This study focuses on the historical city centre of Cambridge (see Figure \ref{fig 2.1-2}), a compact area characterised by high cycling uptake, constrained spatial conditions, and layered morphological complexity. Defined by medieval street grids, narrow corridors, and mixed-use frontages, the area provides a distinctive setting for examining how micro-scale spatial features shape cycling experience \citep{carse_factors_2013, aldred_outside_2010}. The analysis targets 116 discrete street segments within this core, allowing local design features to be assessed under relatively consistent contextual conditions. Street typologies were defined at the segment level by adapting classifications from previous cycling studies \citep{houde_ride_2018, nazemi_studying_2021, noland_understanding_2023, teschke_proximity_2017, zhang_measuring_2022} to Cambridge's street morphology and modal allocation.

\subsubsection{Extraction of indicators}
To provide a controlled and typologically representative visual basis for perception assessment, we developed the Cambridge Cycling Experience Video Dataset (CCEVD), a first-person video corpus capturing real-world cycling in Cambridge's historical core. The dataset was designed for spatial coverage, typological diversity, and consistent recording conditions. Data were collected from 21 to 28 April 2025 during the University of Cambridge Easter break to reduce pedestrian and vehicle traffic, with all recordings conducted in daylight and outside peak commuting hours (09:30 to 11:30 and 13:00 to 17:00). A GoPro Hero 10 was mounted on the handlebar of a single bicycle and operated by the same cyclist, with the camera aligned to a typical eye-level view without obstruction. Each segment was ridden at a steady pace, and footage with forced stoppages such as traffic lights and pedestrian crossings was excluded to preserve continuity of spatial experience. For the perception survey, one uninterrupted 20 s clip was extracted from each segment to reduce cognitive fatigue while retaining contextual clarity.

\begin{figure}
    \centering
    \includegraphics[width= 1\linewidth]{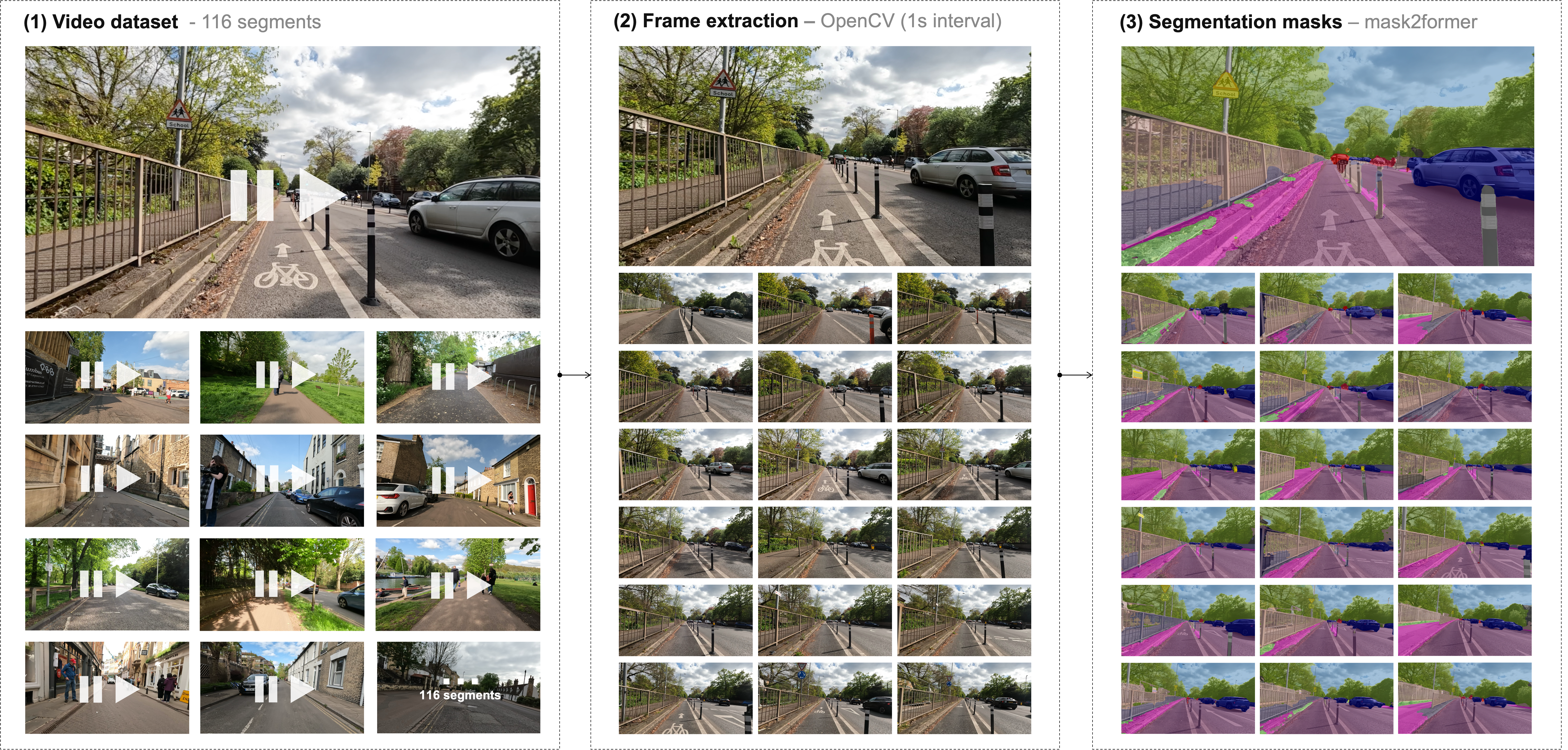} 
    \caption{Cambridge Cycling Experience Video Dataset: quantifying machine-vision streetscape features using OpenCV and semantic segmentation models.}
    \label{fig 3.3.1-2} 
\end{figure}

Segmentation was performed using ZenSVI, an open-source platform for geospatial visual analysis \citep{ito_zensvi_2025}. This implementation employed Mask2Former, a transformer-based semantic segmentation architecture pretrained on Cityscapes \citep{cheng_masked-attention_2022}. The model categorises pixels into 19 urban feature classes, including vegetation, road, sidewalk, and building, and generates pixel-wise class distributions for each frame (see Figure \ref{fig 3.3.1-2}). The mapping between semantic classes and the final machine-vision streetscape indicators is provided in Supplementary Material Section D, Table S5.

As Cityscapes was developed for general urban contexts, a targeted manual visual audit was conducted to assess the transferability of the segmentation outputs to Cambridge's historical core. One frame from each of the 116 street segments was inspected for overall segmentation quality and indicator-relevant errors, and classified as acceptable when no visually apparent error affected the semantic classes used to derive the six machine-vision streetscape indicators. Of the 116 audited frames, 113 met this criterion. The remaining three showed minor errors associated with lake or river surfaces, which are not represented as a separate class in the Cityscapes taxonomy. As water was not included in the operational definitions of the six indicators, these errors were considered unlikely to substantially affect the derived streetscape measures.

Six visual indicators were derived from class proportions: greenness, openness, enclosure, non-motorised lane proportion, imageability, and building continuity. These metrics were computed using defined pixel-class combinations and averaged across all frames for each segment. Full formulas are provided in Supplementary Material Section A, Table S1. Spatial data were sourced from OpenStreetMap (OSM) and the Shuttle Radar Topography Mission (SRTM) Digital Elevation Model (DEM). OSM layers, including buildings, road networks, land-use parcels, POIs, and speed limits, were queried using the QuickOSM plugin in QGIS. DEM data were clipped to the Cambridge study area and reprojected to the British National Grid (EPSG:27700) to ensure alignment across datasets.

\subsubsection{Perception survey}

To capture subjective impressions of the street environment under controlled visual conditions, a video-based perception survey was conducted between 20 and 23 May 2025. Participant demographics and cycling habits are summarised in Supplementary Material Section C, Table S4. Participants viewed 20 s clips extracted from the full-length recordings and evaluated five experiential dimensions: perceived safety, comfort, quality of cycling experience, pleasure, and arousal. This structure incorporated both evaluative and affective components to reflect the complexity of cycling experience.

A Balanced Incomplete Block Design (BIBD) was used to assign 116 clips across participants. Each segment received at least 12 independent ratings \citep{gu_designing_2025}, and each participant rated 20 unique clips. Following the BIBD equation $b \times k = r \times v$, a minimum of 70 participants was required. Stratified sampling ensured that each participant encountered at least five street typologies, while a Python-based allocation script optimised clip distribution and avoided overrepresentation. Minor post-adjustments allowed a small number of segments to receive 13 ratings to complete assignment sets. The BIBD parameters are summarised in Supplementary Material Section C, Table S3.

The survey was administered via Qualtrics using preassigned IDs to deliver individualised clip sets. Perceived safety, comfort, and quality were rated on 5-point Likert scales, while pleasure and arousal were measured using the Affect Grid \citep{russell_affect_1989}. Responses were aggregated at the segment level by averaging participant ratings within each segment, producing five continuous experience indicators for subsequent modelling and typology-based analysis. Full definitions of these indicators are provided in Supplementary Material Section A, Table S1.

To extend experiential assessment beyond the surveyed clips, a Random Forest regression model was trained using semantic segmentation outputs as predictors and aggregated experience indicators as target labels. The trained model was applied to the full video dataset at one-second intervals, with predictions averaged within each segment to generate a secondary set of experience estimates. These predicted values were not used for statistical inference, but informed the composite bikeability index and typology-specific design scenarios.

Bicycle-mounted video recording and the online perception survey were approved by the institutional ethics committee (Ethics ID: AHA/PG/ST24100019). Identifiable faces and personal features were blurred during processing, and all data were stored in accordance with GDPR requirements. All participants provided informed consent before taking part in the study.

\subsubsection{Expert interview}

To complement the quantitative analysis with applied planning perspectives, an expert interview was conducted on 5 June 2025 with Josh Grantham, Infrastructure Campaigner at Cambridge Cycling Campaign (Camcycle). It assessed the practical relevance of the proposed Bikeability Index and typology-specific design scenarios under Cambridge's planning, political, and infrastructural constraints, with particular attention to historically protected areas and perception-oriented street-level adjustments. The expert interview received separate ethics approval from the same committee (Ethics ID: AHA/PG/ST24100039).

\subsection{Statistical analysis}

Two hypotheses were set a priori:

\begin{enumerate}
    \item Openness, greenness, and non-motorised lane proportion are positively associated with perceived cycling experience, including perceived safety, comfort, and affective responses.
    \item Enclosure, imageability, and building continuity show non-linear effects, with moderate levels associated with more positive cycling experience and excessive levels potentially related to discomfort or higher arousal.
\end{enumerate}

\subsubsection{Elastic net regression model}

Elastic Net regression assessed linear relationships between spatial features and five perceptual outcomes: perceived safety, comfort, quality, pleasure, and arousal. Combining Lasso and Ridge penalties, it is suited to intercorrelated predictors and supports coefficient shrinkage, variable selection, and model stability \citep{zou_regularization_2005}. Parameter tuning used ten-fold cross-validation. Bootstrapped standard errors ($n = 1000$) assessed coefficient stability, and residual diagnostics checked model assumptions \citep{stine_introduction_1989, mackinnon_confidence_2004, diep_ufboot2_2018}.

\subsubsection{Random forest + SHAP}

Random Forest regression explored non-linear relationships and potential interactions through bootstrapped decision trees \citep{breiman_random_2001}. Separate models were trained for the five perceptual outcomes using the same predictor set as the Elastic Net models, with a 70:30 train/test split and tuned hyperparameters. SHAP (Shapley Additive Explanations) attributed prediction outputs to individual feature contributions \citep{li_extracting_2022}. Global summary plots ranked feature influence, while dependence plots illustrated effect shapes across value ranges.

\subsubsection{Kruskal-Wallis H-test}

Kruskal-Wallis tests assessed whether perceptual responses differed across street typologies and participant familiarity groups. This non-parametric method is appropriate for ordinal response variables and avoids assumptions of normality \citep{kruskal_use_1952, breslow_generalized_1970}. Each perceptual outcome was tested across six street typologies, with Dunn's post-hoc tests and Bonferroni correction applied where significant differences emerged. Additional tests compared ratings between participants with different levels of familiarity with Cambridge.

\subsection{Bikeability index construction}

\begin{figure}
    \centering
    \includegraphics[width= 1\linewidth]{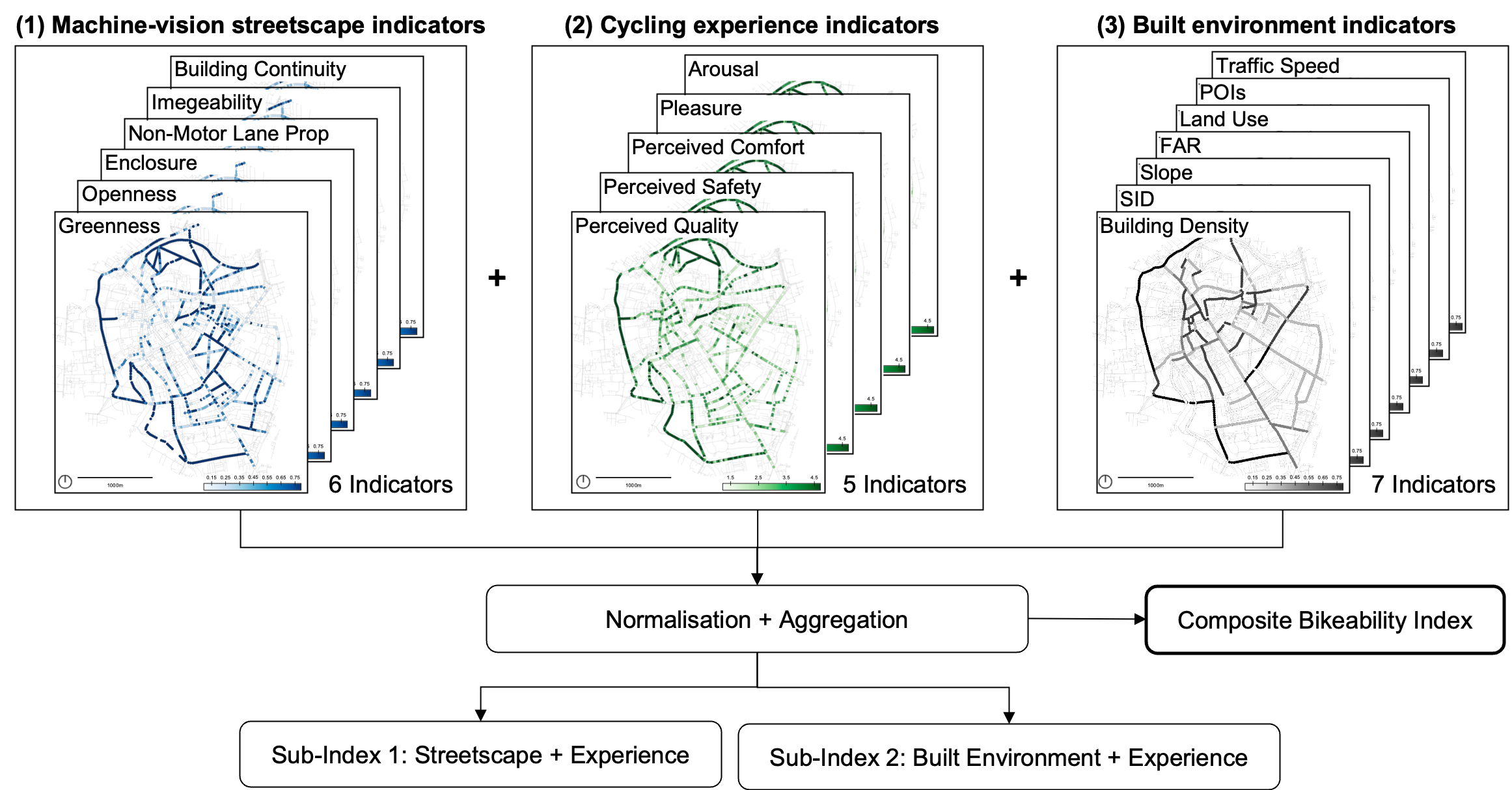} 
    \caption{Indicator composition and Bikeability Index construction.
    \newline The index integrates 18 segment-level indicators from three domains: machine-vision streetscape indicators, cycling experience indicators, and built environment indicators. The machine-vision streetscape domain includes greenness, openness, enclosure, imageability, non-motorised lane proportion, and building continuity. The cycling experience domain includes perceived quality, perceived safety, perceived comfort, pleasure, and arousal. The built environment domain includes building density, street intersection density, slope, floor area ratio, land use, points of interest, and traffic speed. All indicators were normalised and aggregated within their respective domains before aggregation into the composite Bikeability Index. Two diagnostic sub-indices were also constructed: Sub-Index 1 combines streetscape and cycling experience indicators, while Sub-Index 2 combines built environment and cycling experience indicators.}
    \label{fig 4.4-0} 
\end{figure}

As summarised in Figure \ref{fig 4.4-0}, this study evaluates 18 indicators across three categories: machine-vision streetscape features, cycling experience indicators, and built environment attributes. All indicators were computed at the segment level for 116 street segments and rescaled to a 0-1 range using min-max normalisation:

\begin{equation}
x^{\prime}_i = \frac{x_i - \min(x)}{\max(x) - \min(x)}
\end{equation}

Within each category, the normalised indicators were averaged to derive three sub-indices representing streetscape perception ($x^{\prime}_{\text{SP}}$), cycling experience ($x^{\prime}_{\text{CE}}$), and built environment conditions ($x^{\prime}_{\text{BE}}$). These sub-indices were again normalised before final aggregation. The composite Bikeability Index was calculated as the unweighted mean of the three sub-indices and scaled to a 0-100 range:

\begin{equation}
\text{Bikeability Index} = \left( \frac{x^{\prime}_{\text{SP}} + x^{\prime}_{\text{CE}} + x^{\prime}_{\text{BE}}}{3} \right) \cdot 100
\end{equation}

All indicators were retained in their observed orientation after min-max normalisation. The index was designed to capture the combined configuration of spatial, perceptual, and built-environment conditions without assuming a uniform marginal direction for each variable. A directionality check across the five perceptual outcomes identified no indicator with consistently negative associations, providing no empirical basis for systematic reverse coding.

Equal weighting was adopted to avoid imposing subjective priorities among the three domains and to reflect their conceptual complementarity. To test weighting sensitivity, the baseline equal-domain index was compared with equal indicator-level, SHAP-derived, and ElasticNet-derived alternatives. Robustness was evaluated across the 116 segments using Spearman rank correlation, top-20\% overlap, and mean absolute rank change.

All index scores were mapped in QGIS using graduated symbology and overlaid on street typology layers. Two variant indices were also constructed: a micro-based index combining $x^{\prime}_{\text{SP}}$ and $x^{\prime}_{\text{CE}}$ to emphasise street-level perceptual conditions, and a macro-based index combining $x^{\prime}_{\text{BE}}$ and $x^{\prime}_{\text{CE}}$ to highlight broader spatial structure.

\section{Results}
\subsection{Correlational analysis}

Before examining environmental associations, Spearman correlation analysis showed strong relationships among perceived quality, safety, and comfort ($\rho = 0.91$--$0.95$, $p < 0.001$), while arousal was positively correlated with these evaluative dimensions ($\rho = 0.69$--$0.77$, $p < 0.001$). Pleasure showed only weak and non-significant associations with the other outcomes ($|\rho| < 0.14$), indicating related but non-identical perceptual responses across the five measures.

Environmental characteristics were systematically associated with perceived cycling experience across the 116 street segments. Linear relationships were estimated using Elastic Net regression, while Random Forest models with SHAP interpretation were used to examine non-linear patterns and feature interactions. Across outcomes, machine-vision streetscape indicators showed more substantial and varied associations than built environment variables, which served primarily as contextual controls. The five perceptual outcomes were modelled separately: perceived quality, safety, comfort, pleasure, and arousal. Several features showed limited or context-dependent associations, but the main patterns broadly aligned with the initial hypotheses. Elastic Net coefficients are reported in Supplementary Material Section B, Table S2.

\begin{figure}
    \centering
    \includegraphics[width= 1\linewidth]{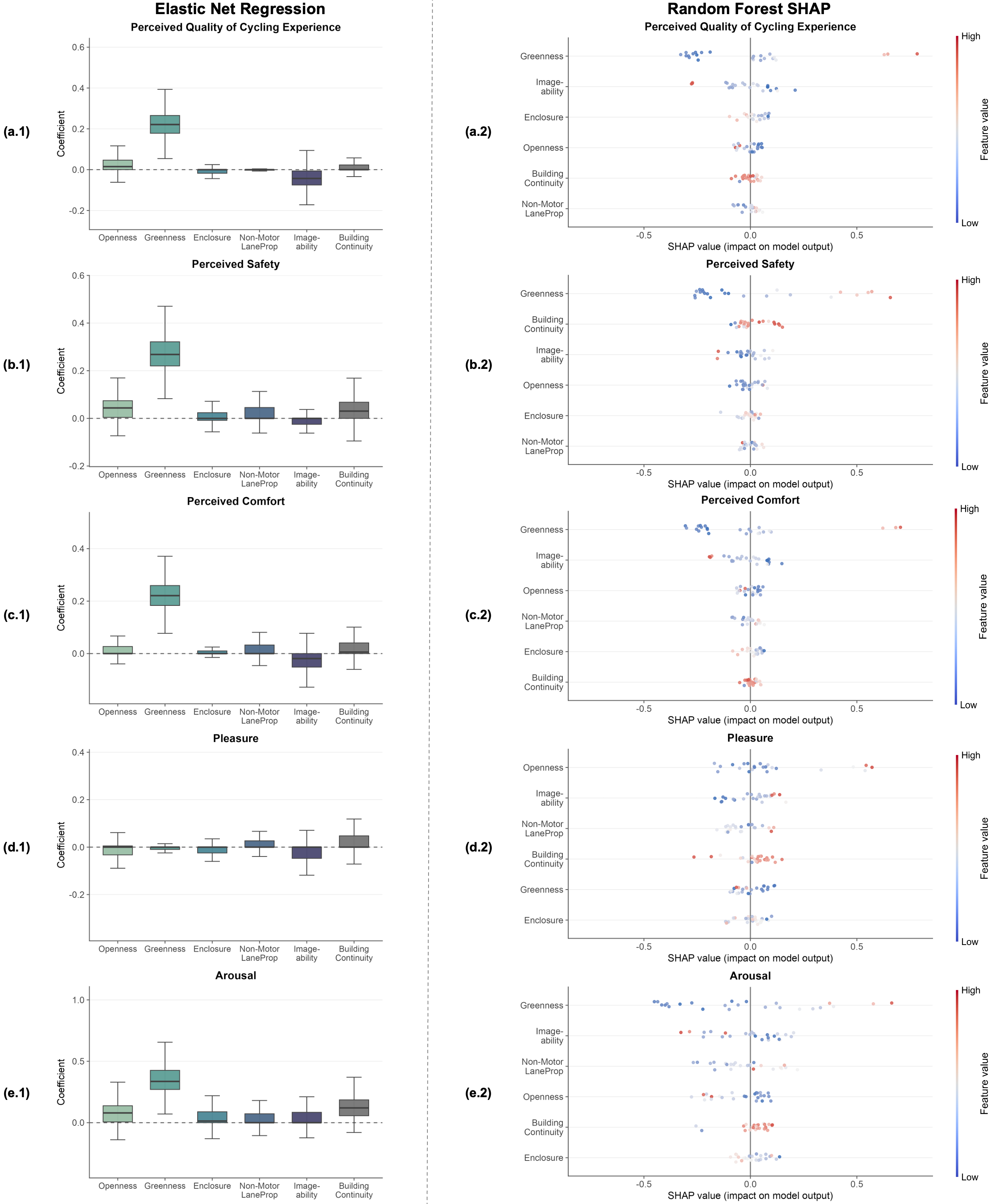} 
    \caption{Correlational analysis.
    \newline The left column shows the distributions of bootstrapped standardized coefficients from Elastic Net regression, with the dashed horizontal line marking zero coefficient. The right column shows SHAP values from Random Forest models, where each point represents a held-out street segment and colour indicates the scaled feature value from low to high. Positive and negative SHAP values indicate whether a feature contributes to higher or lower predicted outcome scores. Rows correspond to perceived quality of cycling experience, perceived safety, perceived comfort, pleasure and arousal.}
    \label{fig 4.2} 
\end{figure}

Greenness showed the strongest and most stable associations across the five outcomes (see Figure \ref{fig 4.2}). It achieved the highest Elastic Net coefficients for safety, arousal, comfort, and quality, and its SHAP values exceeded 0.6. This pattern was visible across typologies, suggesting that vegetation was related to both affective engagement and perceptual clarity. Its role may reflect visual softness and depth cues that support positive evaluations of cycling experience.

Openness and non-motorised lane proportion formed a secondary tier of contributors, both associated with spatial clarity. Openness was positively associated with safety and comfort (see Figure \ref{fig 4.2}.b.1 and \ref{fig 4.2}.c.1), although SHAP values remained modest. Non-motorised lane proportion showed more localised associations. Its SHAP values peaked above 0.4 in select segments, especially for quality and pleasure (see Figure \ref{fig 4.2}.a.2 and Figure \ref{fig 4.2}.d.2), indicating that its perceptual salience corresponded to visible clarity and uninterrupted demarcation. Clearly defined cycle lanes were associated with greater comfort and control, whereas ambiguous or obstructed lanes showed weaker effects.

More complex patterns emerged for imageability, enclosure, and building continuity. Imageability ranked among the top three SHAP contributors for all perceptual outcomes, despite producing negative or negligible coefficients in the Elastic Net models. Its SHAP range exceeded $\pm 0.4$ in safety, pleasure, and arousal, and high imageability observations often appeared on the negative side of the SHAP distribution for evaluative outcomes (see Figure \ref{fig 4.2}.a.2, Figure \ref{fig 4.2}.d.2, and Figure \ref{fig 4.2}.e.2). This pattern indicates a non-linear and context-dependent relationship, although the SHAP summary plots do not establish a specific inflection point. Enclosure showed similarly bifurcated associations: walled boundaries in low-traffic areas corresponded to higher comfort, while compressed edges in shared lanes were associated with higher arousal and lower comfort. Building continuity was most salient for safety, with SHAP values above 0.5 in segments characterised by uninterrupted façades (see Figure \ref{fig 4.2}.a.2), suggesting that continuous frontage may support predictability and lower perceived exposure.

The results indicate two broad patterns. Greenness and openness showed comparatively stable positive associations across perceptual outcomes. Imageability and building continuity were more dependent on spatial configuration, with associations shaped by local morphological composition and movement visibility.

The explanatory power of the models places these associations in context. The Elastic Net regressions produced R$^2$ values between 0.36 and 0.38 for perceived safety, comfort, and quality. These values are comparable to or above thresholds commonly reported in studies using similar machine-vision spatial features \citep{qin_investigating_2024, nelson_bicycle_2023, kawshalya_impact_2022}, but remain moderate, reflecting the complexity of perceptual responses. The typology-specific comparisons in the next section examine this contextual variation in greater detail.

\FloatBarrier

\clearpage
\newpage
\subsection{Comparative analysis}

\subsubsection{Comparisons between street typologies}

\begin{figure}
    \centering
    \includegraphics[width= 1\linewidth]{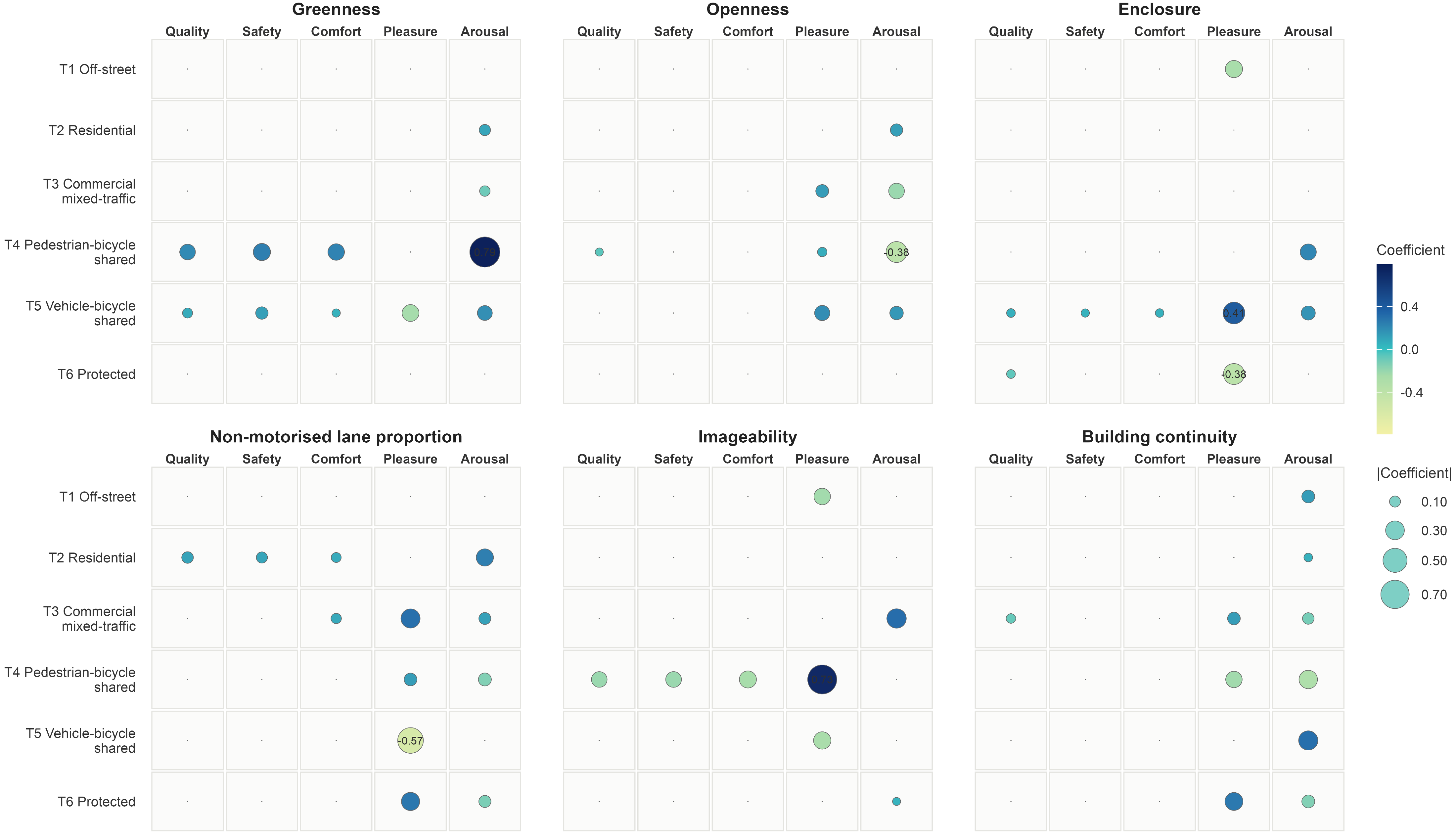} 
    \caption{Typology-specific feature associations across perceived cycling outcomes.
    \newline Each panel represents one machine-vision streetscape feature. Rows represent the six street typologies. Columns show the five perceived cycling outcomes. Bubble size indicates the absolute Elastic Net coefficient magnitude. Bubble colour indicates the coefficient value on a shared scale. Near-zero coefficients are left unmarked to reduce visual noise.}
    \label{fig 4.3.1} 
\end{figure}

Typology-sensitive modelling provides a diagnostic lens for interpreting the contingent nature of environmental perception. While the general correlational analysis in Section 4.1 offered a composite view across all street segments, this subsection isolates within-typology variation through separate Elastic Net regressions for each street type. In parallel, Kruskal-Wallis H-tests examined whether mean values of perceptual outcomes and machine-vision streetscape features differed significantly across typologies. As shown in Table~\ref{tab3.2.1}, all five perceptual dimensions except pleasure exhibited statistically significant differences ($p < 0.01$), as did all streetscape features except enclosure. These results indicate that street typology structured both spatial configuration and perceptual response. Pleasure and arousal, which showed limited explanatory power in the overall models, presented stronger and more varied feature associations when disaggregated by typology. This suggests greater sensitivity to local spatial context.

Within-typology regressions showed that the relevance of individual predictors shifted across street types (see Figure~\ref{fig 4.3.1}). Reorganising the coefficients by streetscape feature clarifies where these shifts were concentrated. Greenness remained important, but its associations were uneven across typologies. In Pedestrian-Bicycle Shared Lanes, greenness showed positive coefficients for arousal ($0.79$) and safety ($0.25$). It also showed a positive coefficient for comfort ($0.23$), but remained weak in Off-Street Bicycle Lanes. Non-motorised lane proportion was positively associated with arousal in Residential Street Bicycle Lanes ($0.25$) and with pleasure in Commercial Mixed-Traffic Lanes ($0.31$). It was also positive for pleasure in Protected Bicycle Lanes ($0.28$), but negative in Vehicle-Bicycle Shared Lanes ($-0.57$). Imageability showed a strong positive coefficient for pleasure in Pedestrian-Bicycle Shared Lanes ($0.73$), but negative coefficients in Protected Bicycle Lanes ($-0.38$) and Off-Street Bicycle Lanes ($-0.23$).

The feature-specific patterns were most visible for pleasure and arousal. Pedestrian-Bicycle Shared Lanes showed the strongest positive associations for greenness and imageability, alongside negative coefficients for openness ($-0.38$) and building continuity ($-0.29$). Vehicle-Bicycle Shared Lanes showed the strongest negative coefficient for non-motorised lane proportion ($-0.57$), with a positive coefficient for enclosure ($0.41$). Residential Street Bicycle Lanes showed positive coefficients for non-motorised lane proportion across arousal ($0.25$), quality ($0.11$), and safety ($0.10$). In Commercial Mixed-Traffic Lanes, imageability was associated with arousal ($0.32$), while non-motorised lane proportion was associated with pleasure ($0.31$). Protected Bicycle Lanes showed positive pleasure coefficients for non-motorised lane proportion ($0.28$) and building continuity ($0.27$), but a negative coefficient for enclosure ($-0.38$). Off-Street Bicycle Lanes showed weaker associations overall, with negative pleasure coefficients for enclosure ($-0.25$) and imageability ($-0.23$). These results indicate that typology structured the direction and strength of feature associations, without implying that typology alone determined perceived cycling experience. These typology-specific associations therefore informed the subsequent design implementation stage by indicating which streetscape features were most relevant within each street type.

\begin{table}
\centering
\renewcommand{\arraystretch}{1.3}
\footnotesize
\begin{tabular}{lcc}
\toprule
\textbf{Variable} & \textbf{H-statistic} & \textbf{p-value} \\
\midrule
Greenness & 51.88 & \textbf{$<$ 0.001}\textbf{***} \\
Openness & 17.17 & \textbf{$<$ 0.01}\textbf{**} \\
Enclosure & 8.35 & 0.138 \\
Non-Motorised Lane Proportion & 36.88 & \textbf{$<$ 0.001}\textbf{***} \\
Imageability & 32.02 & \textbf{$<$ 0.001}\textbf{***} \\
Building Continuity & 34.12 & \textbf{$<$ 0.001}\textbf{***} \\
Perceived Quality of Cycling Experience & 38.05 & \textbf{$<$ 0.001}\textbf{***} \\
Perceived Safety & 30.94 & \textbf{$<$ 0.001}\textbf{***} \\
Perceived Comfort & 35.74 & \textbf{$<$ 0.001}\textbf{***} \\
Pleasure & 5.01 & 0.415 \\
Arousal & 16.25 & \textbf{$<$ 0.01}\textbf{**} \\
\bottomrule
\end{tabular}
\caption{Kruskal-Wallis H-test results for street typology comparisons.
\newline Significance levels: * $p < 0.05$, ** $p < 0.01$, *** $p < 0.001$. H-statistics are derived from Kruskal-Wallis non-parametric rank tests across six typologies.}
\label{tab3.2.1}
\end{table}

\subsubsection{Comparisons between city familiarities}

\begin{table}
\centering
\renewcommand{\arraystretch}{1.3}
\footnotesize
\begin{tabular}{lcc}
\toprule
\textbf{Variable} & \textbf{H-statistic} & \textbf{p-value} \\
\midrule
Perceived Quality of Cycling Experience & 14.33 & \textbf{$<$ 0.001}\textbf{***} \\
Perceived Safety & 6.03 & \textbf{$<$ 0.05}\textbf{*} \\
Perceived Comfort & 9.32 & \textbf{$<$ 0.01}\textbf{**} \\
Pleasure & 0.19 & 0.69 \\
Arousal & 0.015 & 0.97 \\
\bottomrule
\end{tabular}
\caption{Kruskal-Wallis H-test results for city familiarity comparisons.
\newline Significance levels: * $p < 0.05$, ** $p < 0.01$, *** $p < 0.001$. H-statistics reflect differences between familiar and unfamiliar participants.}
\label{tab:kruskal_familiarity}
\end{table}

Participant familiarity with Cambridge was associated with differences in evaluative perceptions, with no corresponding differences in affective responses (see Table \ref{tab:kruskal_familiarity}). Kruskal-Wallis H-tests showed significant differences in perceived quality ($H$ = 14.33, $p$ < 0.001), comfort ($H$ = 9.32, $p$ < 0.01), and safety ($H$ = 6.03, $p$ < 0.05), with familiar participants generally providing more positive ratings. This pattern suggests that prior spatial knowledge may support perceived clarity and ease of navigation, particularly in structurally legible environments.

Pleasure ($p$ = 0.69) and arousal ($p$ = 0.97) showed no significant differences between familiarity groups. These affective responses appeared less related to place memory than to immediate visual stimuli in the video-based task. As cycling is an embodied and continuously attention-demanding activity, short video clips may have limited capacity to capture the affective intensity of uninterrupted riding.

\subsection{Bikeability index}

To move beyond isolated feature-level correlations, this study constructed a Bikeability Index for comparative street-segment diagnosis (see Figure \ref{fig 4.4-0}). Indicators were standardised within their categories and combined using equal weights at both aggregation levels, following the specification described in Section~3.3. Equal weighting was retained to avoid imposing a priori priorities across the three domains, with sensitivity to this choice examined in Section~4.3.1.

\subsubsection{Composite bikeability index}

\begin{figure}
    \centering
    \includegraphics[width= 1\linewidth]{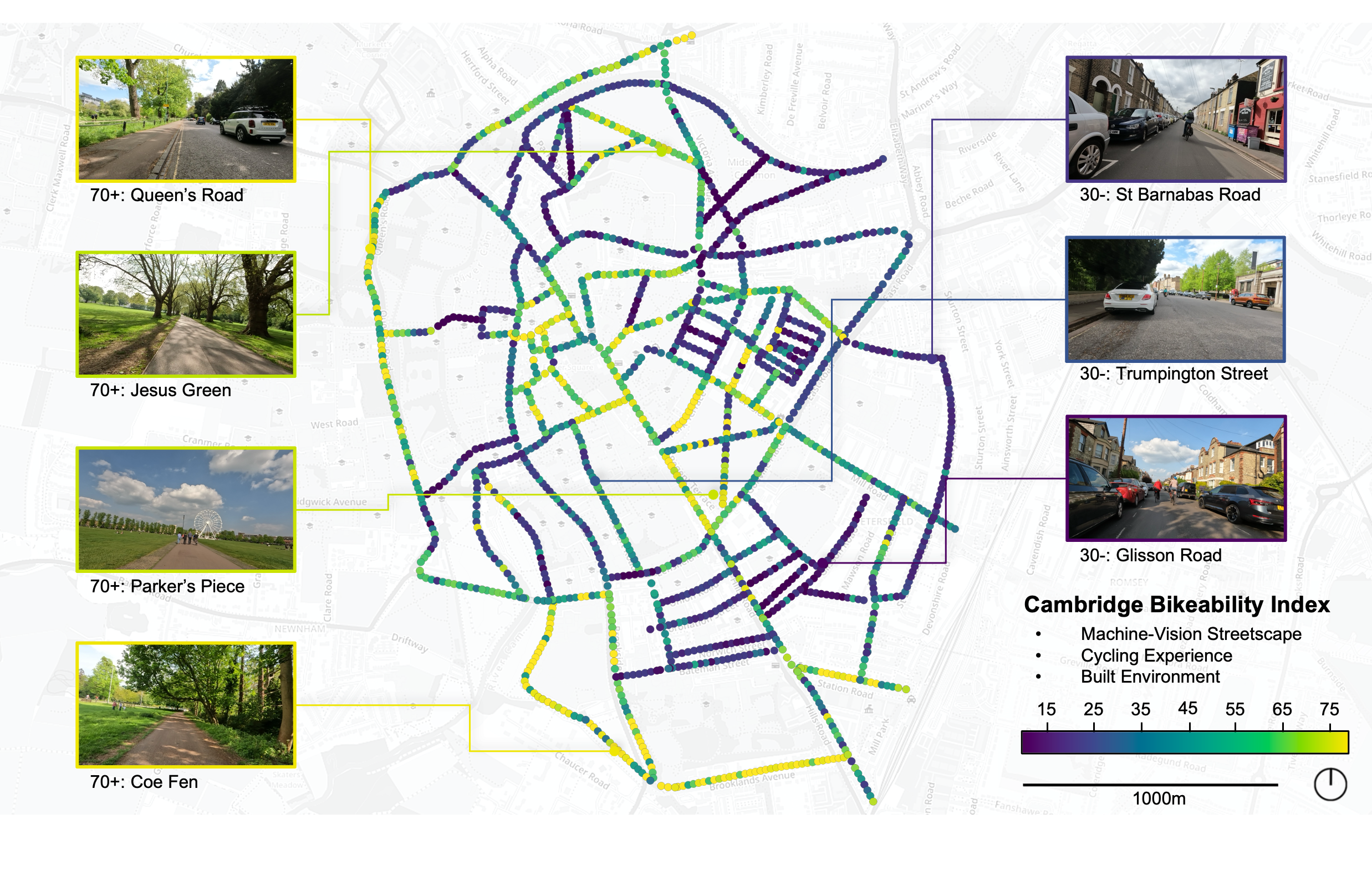} 
    \caption{Cambridge Bikeability Index and representative street segments.
\newline The map shows the composite Bikeability Index across 116 street segments in Cambridge, with higher values indicating more supportive perceived cycling conditions. Representative street-view examples illustrate high-scoring and low-scoring segments, and the colour of each image frame follows the same index scale used in the map.}
    \label{fig 4.4-1} 
\end{figure}

The constructed index shows a wide distribution of perceived bikeability across Cambridge's street network. Scores range from 15.5 to 74.9, with higher values indicating more supportive cycling conditions within the present framework. By combining spatial configuration and perceptual attributes, the index enables direct comparison across street segments and typologies (Figure~\ref{fig 4.4-1}).

The spatial pattern of index values varies substantially across the city centre. High-scoring segments above 70 cluster around green corridors such as \textit{Parker's Piece} and \textit{Jesus Green}, northwest-bound commuting routes, and off-street paths in the southwest. These segments generally combine open or vegetated edges with limited pedestrian interaction and clearer movement cues, corresponding to higher comfort, safety, and pleasure.

Low-scoring segments below 30 are concentrated in high-traffic arterials and constrained residential streets. \textit{Trumpington Street} exemplifies the first condition, where high vehicle volumes, minimal segregation, sparse vegetation, and elevated traffic speeds correspond to lower perceived safety and quality. Dense residential streets form a second low-scoring pattern, where narrow rights-of-way, frequent on-street parking, and continuous building frontages reduce visual openness and imageability.

The index also reveals substantial variation within typologies. Although \textit{off-street} segments tend to score higher and \textit{mixed-traffic} corridors lower, segments within the same class diverge according to local characteristics such as greenness, enclosure, signage legibility, and slope. For example, a \textit{vehicle-bicycle shared lane} along \textit{Queen's Road} scores above 70, reflecting strong vegetative presence, limited pedestrian conflict, and a coherent visual field. By contrast, a segment of the same typology near \textit{Trumpington Street} scores below 30, with fragmented signage, higher intersection complexity, sparse greenery, and increased modal competition.

Sensitivity analysis showed that the overall ranking structure was broadly stable under alternative weighting schemes (Table~\ref{tab4.3.1}). Rank correlations with the equal-domain index remained strong for equal indicator-level weighting ($\rho = 0.945$), SHAP-derived weighting ($\rho = 0.955$), and ElasticNet-derived weighting ($\rho = 0.848$). However, the lower top-20\% overlap under SHAP-derived (75.0\%) and ElasticNet-derived (66.7\%) weighting indicates that high-priority segment selection was more sensitive to weighting assumptions than the overall rank order. The equal-domain index was therefore retained as the main specification.

\begin{table}
\centering
\renewcommand{\arraystretch}{1.3}
\footnotesize
\begin{tabular}{lccc}
\toprule
\textbf{Alternative weighting scheme} & \textbf{Spearman's $\rho$} & \textbf{Top-20\% overlap} & \textbf{Mean absolute rank change} \\
\midrule
Equal indicator-level weighting & 0.945 & 95.8\% & 5.43 \\
SHAP-derived weighting & 0.955 & 75.0\% & 8.14 \\
ElasticNet-derived weighting & 0.848 & 66.7\% & 13.84 \\
\bottomrule
\end{tabular}
\caption{Sensitivity analysis of Bikeability Index rankings under alternative weighting schemes.
\newline Spearman's $\rho$ is calculated against the original equal-domain Bikeability Index. Top-20\% overlap indicates the proportion of segments shared within the highest-ranked 20\% under each specification and the original index. Mean absolute rank change is calculated across 116 street segments.}
\label{tab4.3.1}
\end{table}

\subsubsection{Micro and macro sub-indices}

\begin{figure}[htbp]
    \centering
    \begin{subfigure}[t]{0.32\linewidth}
        \centering
        \includegraphics[width=\linewidth]{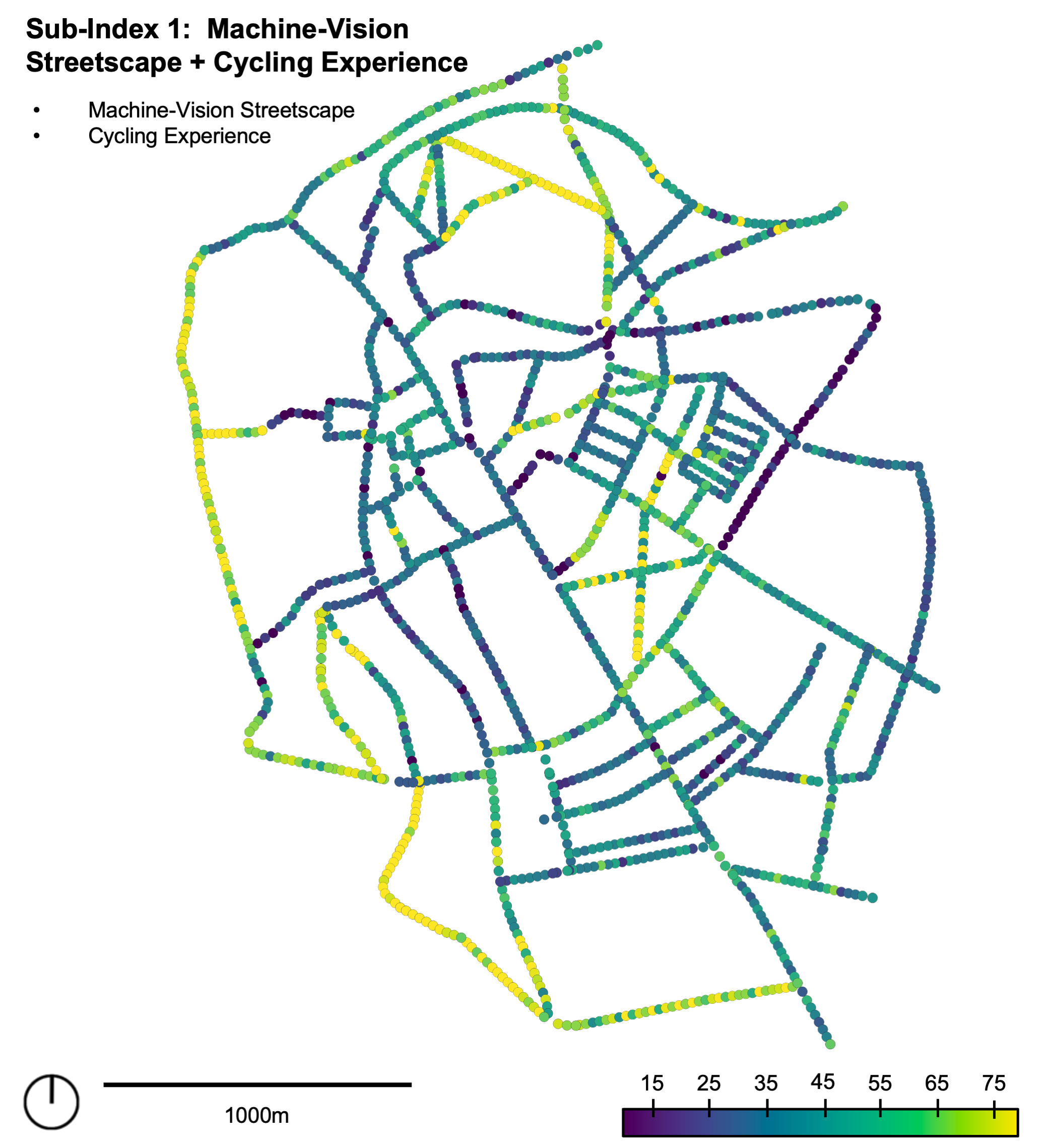}
        \caption{Sub-Index 1: Machine-Vision Streetscape + Cycling Experience}
        \label{fig:subindex1}
    \end{subfigure}
    \hfill
    \begin{subfigure}[t]{0.32\linewidth}
        \centering
        \includegraphics[width=\linewidth]{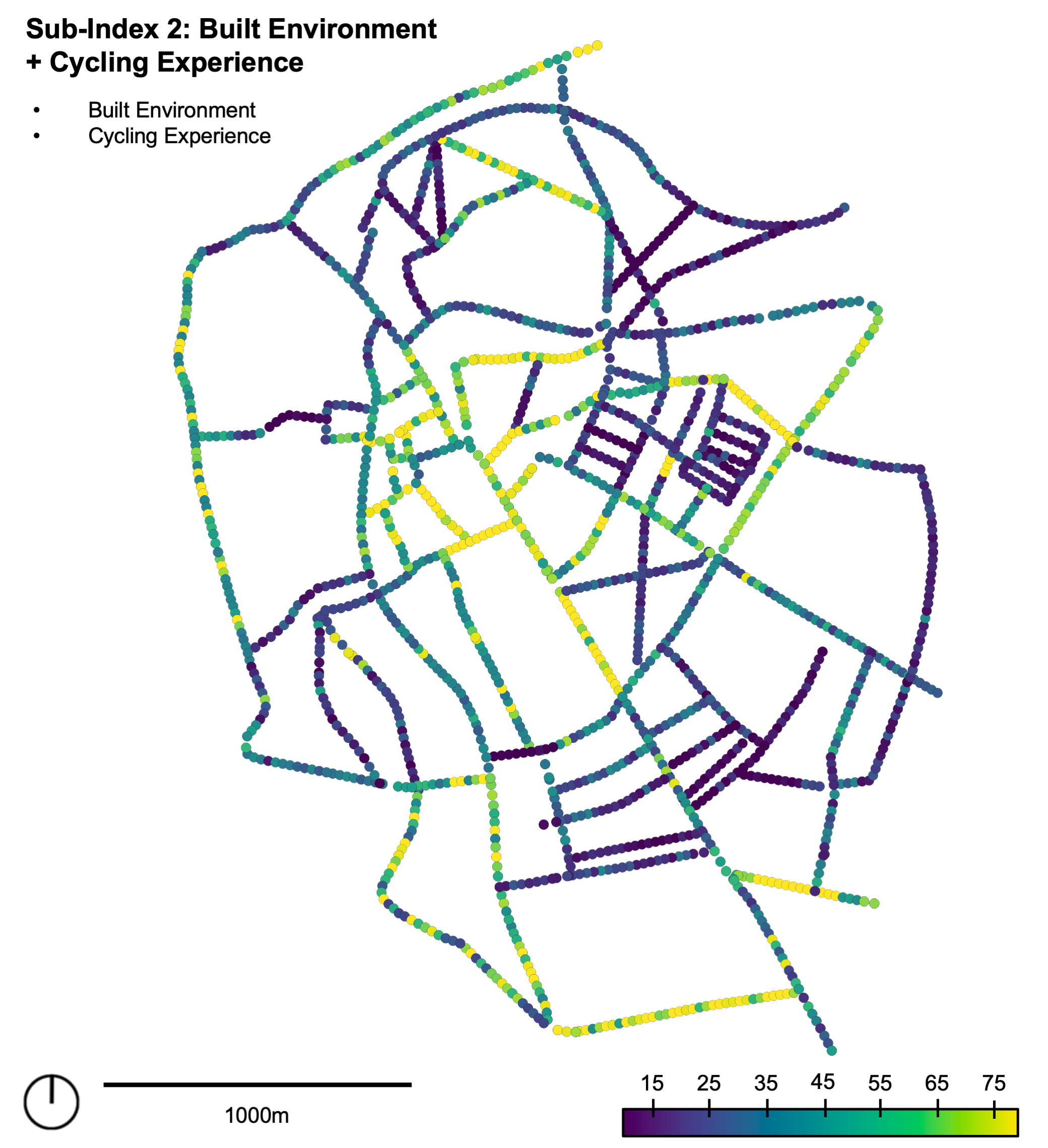}
        \caption{Sub-Index 2: Built Environment + Cycling Experience}
        \label{fig:subindex2}
    \end{subfigure}
    \hfill
    \begin{subfigure}[t]{0.32\linewidth}
        \centering
        \includegraphics[width=\linewidth]{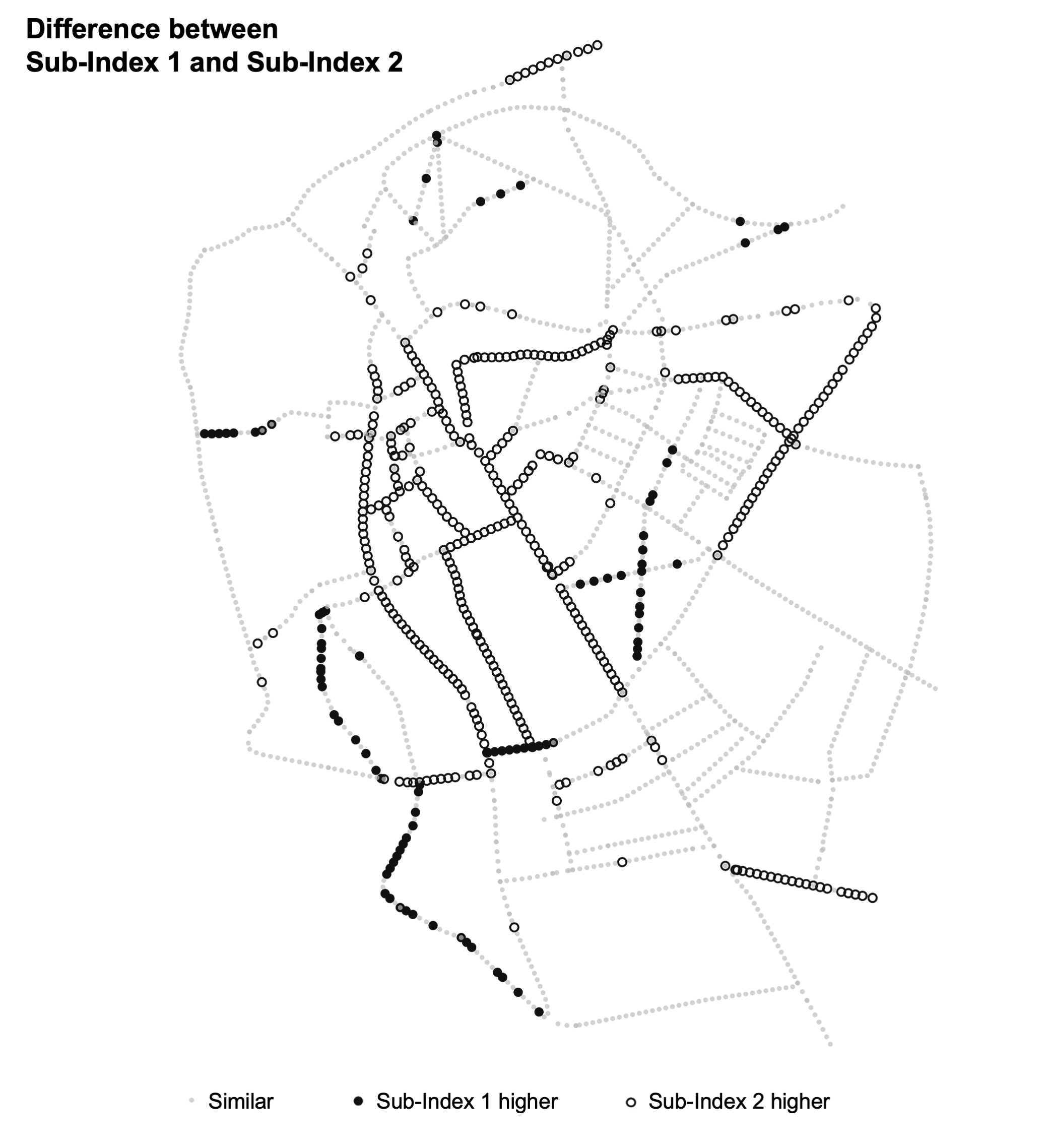}
        \caption{Difference between Sub-Index 1 and Sub-Index 2}
        \label{fig:subindex-difference}
    \end{subfigure}
    \caption{Sub-indices and spatial difference.
    \newline Sub-Index 1 integrates 11 indicators from two domains: (1) Machine-Vision Streetscape (greenness, openness, enclosure, imageability, non-motorised lane proportion, building continuity); (2) Cycling Experience (perceived quality, safety, comfort, pleasure, arousal).
    \newline Sub-Index 2 integrates 12 indicators from: (1) Built Environment (building density, slope, intersection density, floor area ratio, land use mix, POI diversity, traffic speed limit); and (2) Cycling Experience (same 5 experiential indicators).
    \newline The difference map shows frame-level divergence between the two sub-indices, calculated as Sub-Index 1 minus Sub-Index 2. Filled points indicate locations where Sub-Index 1 is higher by more than 15 points, open points indicate locations where Sub-Index 2 is higher by more than 15 points, and light grey points indicate smaller differences.}
    \label{fig:subindices}
\end{figure}

The two sub-indices clarify how street-level perceptual conditions and broader structural conditions contribute differently to bikeability. Sub-Index 1 combines cycling experience with machine-vision streetscape features, while Sub-Index 2 pairs cycling experience with built environment indicators. Both use the same normalisation and weighting, allowing comparison between micro-scale visual conditions and macro-scale morphological patterns, with the difference map identifying where one sub-index exceeds the other (Figure~\ref{fig:subindices}).

Sub-Index 1 shows sharper spatial contrasts than Sub-Index 2, and the difference map makes these local divergences more explicit. Its values range from 18.6 to 73.4 with a standard deviation of 11.2, compared with 22.1 to 71.0 and a standard deviation of 8.6 in Sub-Index 2. Southwest green corridors such as \textit{Coe Fen} and \textit{Parker's Piece}, along with low-traffic shared routes in the northwest, score consistently above 65 in Sub-Index 1. These same segments score 12 to 18 points lower in Sub-Index 2, likely reflecting mono-functional land use and limited amenity density.

By contrast, several central segments attain higher values in Sub-Index 2 above 60. These scores are supported by mixed-use zoning, orthogonal intersections, and regulated speeds, while Sub-Index 1 often falls below 45 where dedicated cycling infrastructure is absent, multimodal traffic is heavier, and visual fields are fragmented. This pattern indicates that structurally legible central streets may still offer a less coherent cycling experience.

Southeast residential areas show the opposite asymmetry. Despite weaker macro attributes, including narrow profiles, limited POI density, and mono-functional use, some segments reach moderate Sub-Index 1 values of 50 to 55. This occurs where vegetation, facade variation, or set-back patterns improve visual openness and reduce perceptual stress, partly offsetting structural limitations.

\subsection{Design implementations}

\begin{figure}
    \centering
    \includegraphics[width= 1\linewidth]{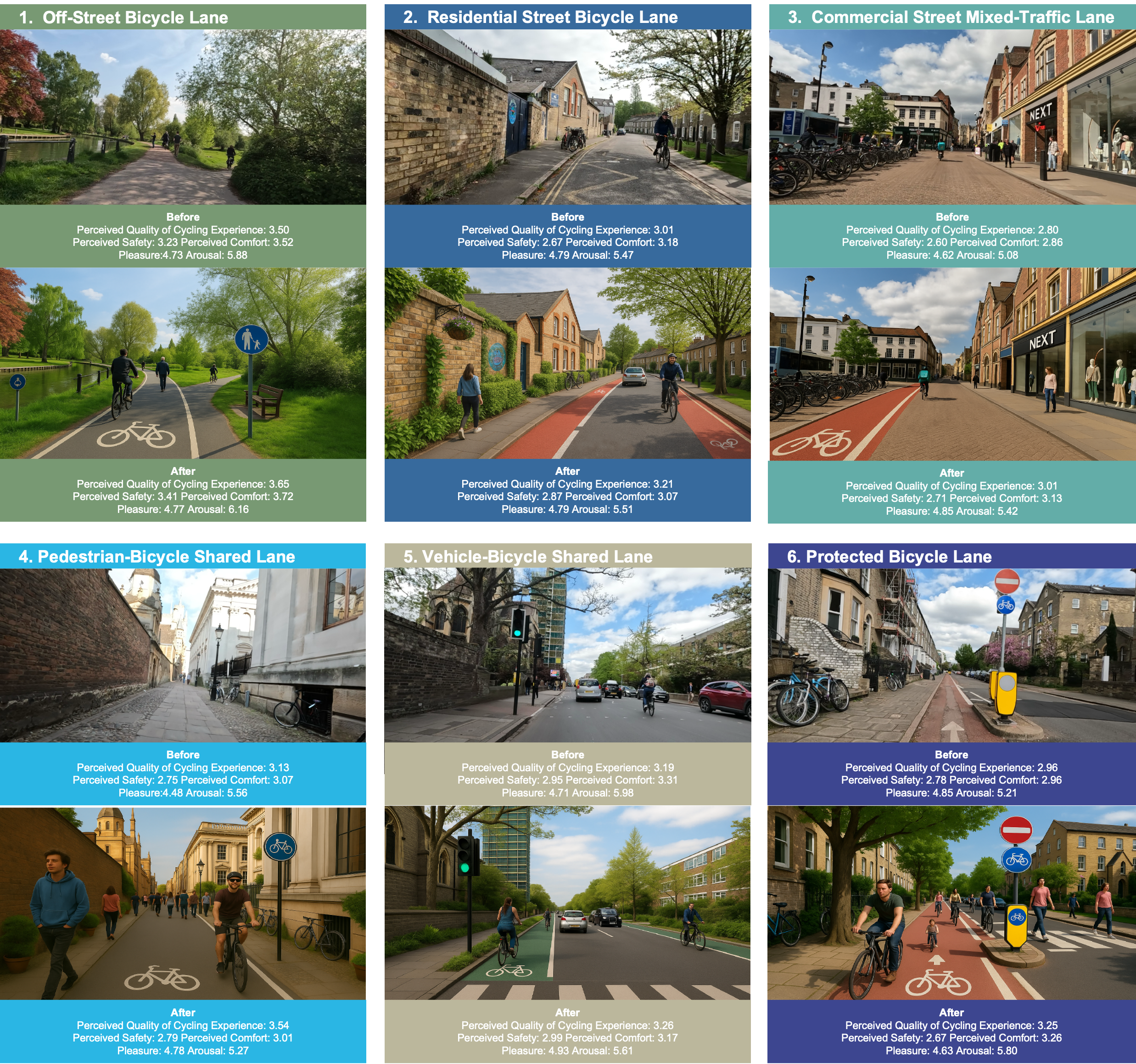} 
    \caption{Model-informed design implementations.
    \newline The figure presents six typology-specific visual scenarios based on the feature associations identified in the statistical analysis. Each scenario compares the original street view with an AI-assisted visual modification and reports Random Forest estimates of perceived quality, safety, comfort, pleasure, and arousal before and after modification.}
\label{fig 4.5-1} 
\end{figure}

The design implementations examine how typology-specific street-level adjustments could respond to the perceptual patterns identified above. Building on the empirical analyses, this section presents AI-assisted visual modifications derived from the main feature associations and typology-specific interpretations (see Figure \ref{fig 4.5-1}). A trained Random Forest was used to estimate perceptual outcomes before and after each visual modification, allowing the scenarios to be compared within the same predictive framework. Because the AI-assisted edits may introduce incidental visual changes beyond the intended street features, the before-after predictions are interpreted at the scenario level and do not isolate the effects of individual design modifications.

The six scenarios varied according to the main perceptual issue identified within each typology. In \textit{Off-Street Bicycle Lanes}, lane markings with gentle separators and reinforced edge vegetation addressed residual path-use ambiguity in an already high-scoring setting. The model estimated small gains in safety ($+0.18$), comfort ($+0.20$), and arousal ($+0.28$).

\textit{Residential Street Bicycle Lanes} required clearer demarcation within narrow streets. Baseline quality and safety were lower (3.01 and 2.67), reflecting obstructive parking and weak lane definition. Red cycling paths with planted buffers and frontage adjustments increased predicted quality and safety by 0.20 and 0.16, while comfort fell slightly in one case ($-0.11$).

In \textit{Commercial Street Mixed-Traffic Lanes}, low baseline safety (2.60) and comfort (2.86) were addressed through stronger visual separation. Raised red-paved bike lanes with edge planting and simplified signage corresponded to estimated gains in safety ($+0.11$) and comfort ($+0.27$), alongside a moderate rise in arousal ($+0.34$).

\textit{Pedestrian-Bicycle Shared Lanes} combined high pleasure and arousal with low safety, indicating visual richness but weak movement hierarchy. Painted icons with subtle paving differentiation and softened walls were used to reduce side clutter. Predictions showed gains in quality ($+0.41$) and safety ($+0.04$), with a small drop in arousal ($-0.29$).

\textit{Vehicle-Bicycle Shared Lanes} showed the highest baseline arousal (5.98), consistent with heavy traffic, obstructed sightlines, and minimal separation. Green cycle lanes with vegetated buffers and visual speed cues corresponded to lower predicted arousal ($-0.37$), higher safety ($+0.24$), and higher pleasure ($+0.22$).

\textit{Protected Bicycle Lanes} already performed relatively well for clarity and structure, with average safety and comfort of 2.78 and 2.96. Restrained surface colouring with directional symbols and soft landscaping produced modest estimated gains in quality ($+0.29$), comfort ($+0.30$), and arousal ($+0.52$).

\section{Conclusion and discussion}
\subsection{Results and relevance}

This paper has set out to address the central research question: how and to what extent do specific street-level features, across different street typologies, shape subjective cycling experiences in the historical city centre of Cambridge?

To answer this question, the project developed a perception-led and typology-sensitive framework linking first-person cycling videos, machine-vision streetscape indicators, built environment variables, and survey-based perceptual responses at the street-segment scale. It produced a street-level Bikeability Index and typology-specific design scenarios grounded in statistical analysis.

\subsubsection*{(1) Which perceptual indicators exhibit the most significant influence on subjective cycling experiences?}

Findings from the correlational and machine-learning analyses show that street-level features were associated with perceived cycling experience, but these associations were not uniform across contexts. Greenness and openness showed the most stable positive associations, while imageability, enclosure, and building continuity depended more strongly on street morphology. These results extend prior machine-vision studies that often assume broadly transferable effects \citep{bialkova_how_2022, rui_beyond_2024, zhang_encouraging_2024, ramirez_juarez_cyclists_2023, guo_moderation_2024, ma_measuring_2021, cain_development_2018}. The models achieved average R$^2$ values between 0.30 and 0.40, with stronger fit observed in several typology-specific regressions. This level of explanatory power is higher than most comparable approaches \citep{qin_investigating_2024, nelson_bicycle_2023, kawshalya_impact_2022, ding_how_2025, wang_systematic_2025}, suggesting that first-person streetscape indicators can explain part of the variation in cycling perception while remaining sensitive to local spatial context.

Greenness emerged as the most stable positive predictor across the five perceptual dimensions, especially for perceived comfort and pleasure. This suggests that vegetation may contribute to more positive cycling evaluations, aligning with studies linking eye-level greenness to cycling uptake and positive street perception \citep{wu_does_2020, wang_relationship_2020, lu_associations_2019}. By contrast, other studies report ambiguous or negative effects where head-level vegetation obstructs visibility or appears unmanaged \citep{sun_dockless_2024, ye_visual_2019, ewing_streetscape_2016}. Greenness therefore appears generally supportive, although its perceptual value remains dependent on visibility, maintenance, and spatial fit.

Openness and non-motorised lane proportion were also associated with safety and comfort, but their effects were more context-dependent. Both corresponded to more positive experiences in legible settings such as \textit{Parker's Piece} and \textit{the Fen Causeway}, while their associations weakened in visually or operationally ambiguous morphologies such as \textit{Trumpington Street} and \textit{King's Parade}. Similarly, enclosure, imageability, and building continuity showed no stable association across all settings, with some relationships reversing by spatial configuration. This pattern is consistent with the broader affordance-based view that visible street conditions are interpreted in relation to possible movement \citep{gibson_ecological_1979, gotardi_affordance-based_2024}. These features were more positively related to perception in ordered environments, while being associated with higher arousal or lower comfort in narrow or visually fragmented streets.

Insights from the expert interview further support the importance of spatial legibility. Josh Grantham highlighted clarity and consistency in street-level design, emphasising that \textit{'clearly defined, physically segregated cycle lanes are the most welcomed intervention type, wherever there is space to implement them'}. This observation echoes the model results without treating the statistical associations as causal effects. Results from the comparative analysis further indicate that greenness and imageability retained relevance across typologies, yet shifted with morphological structure.

\subsubsection*{(2) How do different street typologies in Cambridge influence subjective cycling experience?}

With respect to this sub-question, typology-specific regression models, Kruskal-Wallis H-tests, and typology-specific design scenarios were used to examine how street typology shaped the relationship between environmental features and cycling perception in Cambridge's historical city centre. Results indicate that typology helps interpret how street-level features operate differently across spatial settings, beyond its role as a spatial classification. This supports a typology-sensitive approach in which the same feature may acquire different perceptual meanings under different morphological conditions.

In comparative terms, \textit{Off-Street Bicycle Lanes} and \textit{Protected Bicycle Lanes} consistently showed the highest perceptual ratings across the five dimensions. These outcomes were associated with clearer spatial organisation and functional separation, through which generally moderate features became more perceptually salient. For instance, in \textit{Jesus Green} and \textit{Pembroke Street}, clear spatial organisation appeared to strengthen the relevance of enclosure, which remained weak in the general model but became more closely related to comfort and safety within coherent typological settings.

By contrast, shared-road typologies, including \textit{Vehicle-Bicycle Shared Lanes}, \textit{Pedestrian-Bicycle Shared Lanes}, and \textit{Commercial Mixed-Traffic Streets}, produced lower and more inconsistent evaluative scores. In these environments, visual and functional ambiguity appeared to limit the positive role of other perceptual features and bring stress to the foreground. For instance, \textit{Trumpington Street (Vehicle-Bicycle Shared)}, despite partial segregation and moderate greenness, was perceived as the most unsafe and stressful segment in the pilot study. This condition was echoed by Josh Grantham during the expert interview: \textit{'A shared road is structurally confusing. Even with visual greening or some width, it doesn't fix the sense that cyclists have no defined place. Once the stress kicks in, nothing else really registers.'}

Notably, Kruskal-Wallis tests revealed statistically significant differences across typologies for perceived safety, comfort, quality, and arousal, while pleasure did not show a significant typology effect. This pattern suggests that typology was more strongly related to evaluative judgement and activation than to positive affect alone. However, these tests do not isolate traffic volume, time of day, or other unmeasured temporal dynamics, so typology should be interpreted as a spatial context with possible confounding from unmeasured temporal conditions.

The typology-specific scenarios further translated these findings into model-informed design interpretation, a step less commonly developed in previous bikeability studies \citep{noland_understanding_2023, ren_effects_2023, barrero_asking_2022, cai_sidewalk-based_2024, nikiforiadis_investigating_2023}. Drawing on the perceptual patterns identified in the statistical models, each scenario was tailored to the spatial logic of its typology and evaluated through Random Forest before-after estimates. These estimates clarify differentiated intervention priorities under constrained street conditions: shared and mixed-traffic typologies point to clearer modal demarcation and edge management, while protected and off-street lanes mainly require restrained improvements to surface clarity, signage, and maintenance.

\subsubsection*{(3) What spatial patterns emerge from the street-level bikeability index, and how can they inform design interpretation?}

As the integrative outcome of this study, the street-level Bikeability Index offers a spatialised representation of perceived cycling experience across 116 segments in Cambridge's historical centre. By combining 18 indicators from machine-vision streetscape, subjective survey responses, and built environment features, the index translates fragmented street-level conditions into a perception-based assessment. It enables diagnosis of where supportive or problematic cycling conditions cluster, and why segments within the same typology may still perform differently.

The resulting scores suggest that bikeability is shaped by the configuration of local features within each street context. High-scoring segments often combine good visibility, separation, and greenness, supporting perceptual stability and lower navigational stress, as seen in \textit{Queen's Road}, \textit{Jesus Green}, \textit{Parker's Piece}, and \textit{Coe Fen}. By contrast, low-scoring segments such as \textit{St Barnabas Road}, \textit{Trumpington Street}, and \textit{Glisson Road} are marked by fragmented frontages, weak separation, or visual clutter, which may diminish comfort and perceived safety.

The sensitivity analysis further showed that the overall ranking structure remained broadly stable under alternative weighting schemes, although the identification of high-priority segments was more responsive to the weighting specification. This suggests that the index is more robust as a comparative diagnostic measure than as a fixed threshold for segment prioritisation.

The two sub-indices clarify how micro- and macro-scale conditions contribute differently to bikeability. Street-level perceptual conditions often revealed variation that broader infrastructural measures alone would miss, especially in Cambridge's historical core, where large-scale transformation is constrained by narrow rights-of-way, heritage settings, and intense pedestrian activity. These conditions make heritage governance and visitor movement part of the cycling environment, shaping both the feasible design space and the level of visual or modal crowding experienced by cyclists. This interpretation was supported by the expert interview, where Josh described cycling conditions in Cambridge as \textit{'messy narrow streets, stressed to be within cars and tourists'}. While \textit{'more segregation, fewer shared spaces, and just more room'} would be ideal, he noted that \textit{'we don't have space to do that, so we have to make the most of what's already there'}.

At the same time, the index should be interpreted within the limits of the data collection strategy. Josh pointed to subtle disconnections between the index and his lived experience, noting that \textit{Queen's Road} was \textit{'hell during peak hours'} despite being \textit{'like heaven at 6am'}. This divergence may reflect the low-traffic recording period used for the perception survey clips, and indicates that temporal variation remains important for future validation.

Despite these limitations, the index remains a useful diagnostic tool when compared with macro-scale and SVI-based approaches \citep{gao_pedaling_2025, liu_development_2020, battiston_revealing_2023, ito_zensvi_2025, peterson_mapping_2017, ito_translating_2024}. Macro-scale studies often overlook street segments as the immediate units of cycling experience \citep{gao_pedaling_2025, liu_development_2020, battiston_revealing_2023}, while SVI-based studies commonly rely on vehicle-mounted imagery that does not reflect the cyclist's field of view \citep{ito_zensvi_2025, peterson_mapping_2017, ito_translating_2024}. By contrast, this index links first-person video data with subjective perception, offering a case-based framework for examining cycling experience in historically constrained urban environments.

\subsection{Limitations}
To the best of our knowledge, URBAN-SPIN represents a novel, perception-led and typology-based framework for assessing bikeability in historical urban contexts. Although Cambridge was used as a single case study, this bounded scope allowed the framework to be tested through consistent first-person video collection, perception survey design, and segment-level data integration across the historical city centre. The results of the present study are based on the Cambridge dataset, but the data-driven workflow is transferable to other historical urban contexts where large-scale infrastructural change is often limited. The BIBD survey secured 12 independent ratings for each of 116 street segments, supporting segment-level aggregation and modelling. However, segment counts differed across typologies, which may limit statistical power for less-represented street types such as Protected Bicycle Lanes. Larger and more diverse samples would strengthen external validity and better capture heterogeneity in cycling confidence, familiarity, trip purpose, and risk tolerance.

First-person video captures dynamic cues from a cyclist's perspective but limits scalability and automation \citep{guo_multiple_2024, liu_evaluating_2020, lehtonen_evaluating_2016, costa_context_2017}. Although creating such datasets is labour-intensive, Cambridge's compact size and the need for consistent segment-level comparison made the Cambridge Cycling Experience Video Dataset a pragmatic balance between spatial coverage and feasibility. Video-based surveys also cannot fully reproduce embodied and interactive cycling conditions, especially because the recordings were collected by a single cyclist within a fixed time window and therefore could not capture the full variation in seasonality, traffic conditions, and temporal use. By contrast, real-world experiments introduce environmental variability in weather, traffic, and participant behaviour, reducing consistency across participants and limiting scalability \citep{halefom_how_2022, bi_bicycle_2023, bao_understanding_2023, guo_psycho-physiological_2023, mertens_differences_2016}. Therefore, video methods remain valuable for controlled perceptual assessment while their limitations should be considered during research design and result interpretation.

The proposed bikeability index relies on subjective ratings captured through Likert scales and the Affect Grid, which, despite wide use \citep{de_geus_psychosocial_2008, lemieux_how_2009, nazemi_studying_2019}, depend on conscious introspection and post hoc reporting. This approach supports systematic comparison across 116 street segments but may miss subtle, moment-to-moment changes during dynamic movement. Future work should integrate objective measures such as eye-tracking, EEG, and physiological sensing to complement and validate self-reported perceptual responses.

\subsection{Future research directions}

Future research should incorporate objective physiological measures to capture perceptual responses that may be too rapid or implicit for self-report \citep{bretter_emotions_2025, bogacz_modelling_2021, abbas_neuroarchitecture_2024}. Mobile electroencephalography (EEG), galvanic skin response (GSR), and eye-tracking can detect cognitive load and attentional shifts during movement, providing a continuous record of how cyclists respond to changing street conditions. In parallel, immersive technologies such as Virtual Reality (VR) and Mixed Reality (MR) offer a complementary path for manipulating spatial features under controlled conditions while approximating selected aspects of embodied cycling experience \citep{zeuwts_using_2023, zhao_responsive_2020, beirens_which_2024, ghanbari_use_2024}.

A second direction concerns the validation of design scenarios after implementation. Future research should examine how existing design actions have shaped cycling experience over time \citep{wang_relationships_2023, wang_assessing_2024, rerat_build_2024}. Longitudinal assessments can establish whether perceptual effects persist by tracking perceived safety, comfort, and quality before and after specific actions. Small pilot interventions, such as painted lanes or adjusted signage, can also be evaluated through video or VR surveys, or in situ with local authorities using real-time physiological and behavioural data, especially where heritage constraints and visitor activity limit larger infrastructural change.

A further direction is to model cycling perception as a time-resolved process. In cognitive science, this short-term carry-over is formalised as serial dependence, whereby recent stimuli bias ongoing perception toward continuity \citep{lau_visual_2019, gallagher_stimulus_2022, manassi_continuity_2024}. Applied to urban cycling, this suggests a testable possibility: impressions formed in one street segment may influence responses to the next, especially along continuous corridors or repeated morphological conditions. Future research should therefore examine spatial dependence across adjacent segments, using spatial autocorrelation or spatial regression models to test whether perceptual responses remain correlated beyond local feature values. Context-aware multimodal systems and generative agents may eventually support exploratory simulations of perception-informed behaviour, but their use for cycling research would require careful validation against observed behavioural and perceptual data \citep{froehlich_streetreaderai_2025, noyman_travelagent_2025, tan_visual_2025, gath-morad_visibility_2021, adornetto_generative_2025, su_interactive_2025}.

Overall, URBAN-SPIN shows that perceived bikeability is shaped by the interaction between first-person streetscape conditions and street typology. Greenness and openness provided the most consistent perceptual support, while imageability, enclosure, and building continuity depended more strongly on local spatial configuration. By linking cyclist-view video, survey-based perception, and segment-level modelling, the framework offers a transferable way to diagnose cycling conditions in historically constrained urban settings and to translate empirical findings into design-oriented scenarios.










\bibliographystyle{cas-model2-names}

\bibliography{main}



\end{document}